\documentclass[12pt]{article}
\usepackage{epsf, cite, amssymb}
\usepackage{epsfig}

\makeatletter
\renewcommand\section{\@startsection {section}{1}{\z@}%
                                   {-3.5ex \@plus -1ex \@minus -.2ex}
                                   {2.3ex \@plus.2ex}%
                                   {\normalfont\large\bfseries}}
\renewcommand\subsection{\@startsection{subsection}{2}{\z@}%
                                     {-3.25ex\@plus -1ex \@minus -.2ex}%
                                     {1.5ex \@plus .2ex}%
                                     {\normalfont\bfseries}}
\makeatother

\newcommand{\bea}{\begin{eqnarray}}
\newcommand{\eea}{\end{eqnarray}}
\newcommand{\be}{\begin{equation}}
\newcommand{\ee}{\end{equation}}

\newcommand{\hlf}{\frac{1}{2}}

\newcommand{\Z}{{\mathbb Z}}
\newcommand{\R}{{\mathbb R}}
\newcommand{\M}{{\cal M}}

\newcommand{\F}{\Psi}

\newcommand{\com}[2]{{ \left[ #1, #2 \right] }}
\newcommand{\acom}[2]{{ \left\{ #1, #2 \right\} }}

\newcommand{\m}{\mu}

\newcommand{\tr}{{\rm Tr}}
\def\com#1#2{{ \left[ #1, #2 \right] }}
\def\acom#1#2{{ \left\{ #1, #2 \right\} }}
\newcommand{\tq}{\widetilde{q}}
\newcommand{\tp}{\widetilde{p}}
\newcommand{\f}{\psi}
\newcommand{\tf}{\widetilde{\f}}

\newcommand{\Om}[1]{\ensuremath{\mathrm{O}#1^-}}
\newcommand{\Op}[1]{\ensuremath{\mathrm{O}#1^+}}
\newcommand{\Omt}[1]{\ensuremath{\widetilde{\mathrm{O}#1}{}^-}}

\newcommand{\D}[1]{\ensuremath{\mathrm{D}#1}}

\newcommand{\C}[1]{$(\ref{#1})$}

\newcommand{\xs}{\not\!\!X}
\newcommand{\ps}{\not\!\!P}
\newcommand{\dif}{{d}}
\begin{document}
\begin{titlepage}

\begin{center}

\today
\hfill                  hep-th/0205162

\hfill EFI-02-71

\hfill LPTENS-02/27, PAR-LPTHE 02-26

\vskip 2 cm
{\Large \bf Twisting $E8$ Five-Branes}\\
\vskip 1.25 cm
{
Arjan Keurentjes$^a$\footnote{email address:
Arjan.Keurentjes@lpt.ens.fr} and Savdeep Sethi$^b$\footnote{email address:
 sethi@theory.uchicago.edu}}\\
\vskip 0.5cm
{\sl{}
$^a$  LPTHE, Universit\'e Pierre et
Marie Curie, Paris VI, Tour 16,\\ 4 place Jussieu, F-75252 Paris Cedex 05,
France\\
and \\
Laboratoire de Physique Th\'eorique de l'Ecole
Normale Sup\'erieure, \\24 rue Lhomond, F-75231 Paris Cedex 05, France\\
\vskip 0.2cm
$^b$ Enrico Fermi Institute, University of Chicago, Chicago, IL
60637, USA\\
and \\
School of Natural Sciences, Institute for Advanced Study, Princeton, NJ 08540
}

\end{center}

\vskip 1 cm

\begin{abstract}

\baselineskip=18pt

We consider the tensor theory on coincident $E8$ $5$-branes
compactified on $T^3$. Using string theory, we predict that there
must be distinct components in the moduli space of this theory.
We argue that new superconformal field theories are to be found
in these sectors with, for example, global $G_2$ and $F_4$
symmetries. In some cases, twisted $E8$ $5$-branes can be
identified with small instantons in non-simply-laced gauge
groups. This allows us to determine the Higgs branch for the
fixed point theory.

We determine the Coulomb branch by using an M theory dual
description involving partially frozen singularities. Along the
way, we show that a D0-brane binds to two D4-branes, but not to
an $Sp$-type O4-plane (despite the existence of a Higgs branch).
These results are used to check various string/string dualities
for which, in one case (quadruple versus NVS), we present a new
argument. Finally, we describe the construction of new non-BPS
branes as domain walls in various heterotic/type I string
theories.


\end{abstract}

\end{titlepage}

\pagestyle{plain}
\baselineskip=19pt

\section{Introduction}

The $5$-brane of the $E_8\times E_8$ heterotic string can be
viewed as the small size limit of an $E_8$
instanton~\cite{Ganor:1996mu, Seiberg:1996vs}. Supported on the
$5$-brane is a $(1,0)$ tensor theory which is both mysterious and
fascinating. From its origin as the zero size limit of an
instanton, we know that the theory has a Higgs branch
parametrized by $29$ massless hypermultiplets. This branch
describes the $E_8$ instanton together with its moduli.  There is
also a Coulomb branch parametrized by the single scalar of a
$(1,0)$ tensor multiplet. The expectation value of the scalar
determines the position of the $5$-brane in the M theory
direction. At the intersection of these two branches is a
superconformal field theory with $8$ supercharges, and global
$E_8$ symmetry. For $N$ $5$-branes, the structure is similar. The
Higgs branch has $30N-1$ light hypermultiplets, while the Coulomb
branch has $N$ light tensor multiplets.

While little is known about interacting tensor theories, it is
conventional wisdom that when compactified on a torus, these
theories reduce to Yang-Mills theories. Compactifications of
small instanton theories have been studied in~\cite{Ganor:1996xd,
Ganor:1996pc, Intriligator:1999cn}. One of the interesting
properties of Yang-Mills theory, first discussed
in~\cite{'tHooft:1979uj}\ for the case of $SU(N)/\Z_N$, is the
possibility of turning on `non-abelian magnetic flux' on a
$2$-cycle. More generally, for gauge group $G/Z(G)$ where $G$ has
center $Z(G)$, the magnetic flux on a space $\M$ is classified by
$H^2(\M, Z(G))$.

We might then imagine that the choice of flux on $T^3$ can be
studied by first reducing to gauge theory on one circle, and then
studying the possible 't Hooft twists in this gauge theory. Using
string theory, we shall see that this is not true for the $E_8$
$5$-brane: there are new sectors on $T^3$ with no corresponding
gauge theory interpretation. These sectors are distinguished by a
kind of tensor flux analogue of magnetic flux. This is fairly
basic property of these interacting $(1,0)$ theories that we might
hope to understand from first principles.

The effective $2+1$-dimensional physics in these exotic sectors
includes  interacting superconformal field theories with $16$
superconformal charges, and various exotic global symmetries
(listed in table~\ref{table:symmetry}). To each of these new fixed
points labeled by the number of branes $N$, there should
correspond an $AdS_4$ gauged supergravity with $16$
supersymmetries. Establishing the existence of these theories
would provide a beautiful link between the classification of flat
bundles in gauge theory  -- in this case, triples of commuting
connections -- and gauged supergravities.

We determine the Higgs and Coulomb branches for these fixed point
theories in the following way: the structure of the Higgs branch
follows from viewing these branes as small instantons in
non-simply-laced gauge groups.\footnote{A technical remark is in
order: we will study standard Yang-Mills instantons embedded in
higher dimensional theories. The brane fills the dimensions
transverse to the instanton. While $4$-dimensional instantons are
scale invariant because Euclidean Yang-Mills is classically
conformally, this is not true for Yang-Mills theories in other
dimensions. To avoid the instability that makes the instanton
want to shrink, we will completely compactify the spatial
directions transverse to the instanton. Our brane then has finite
volume, and therefore finite mass. Only some of the compact
directions along the brane will play a role in our analysis. In
subsequent discussion, it should be implicitly understood that the
scaling problem is solved this way.} These instantons can be
studied by probing various orientifold $4$-planes with D0-branes.
In this way, we resolve some puzzles in the ADHM construction for
orthogonal groups. We also show that in the case of a pure
\Op{4}-plane which supports no space-time gauge group (but with
D4-branes supports an $Sp$ group), there are still localized
``$Sp(0)$'' instantons. Nevertheless, using $L^2$ index theory
and the theorem of~\cite{Sethi:2000zf}, we show that a bulk
D0-brane does not bind to an \Op{4}-plane even though there is a
Higgs branch. It does, however, bind in a unique way to $2$
D4-branes. Since a D0-brane binds uniquely to $1$
D4-brane~\cite{Sethi:1996kj}, it seems highly likely that it
binds uniquely to any number of D4-branes. These results agree
with the analysis of~\cite{Dorey:2000zq, Dorey:2001ym}, where the
bulk term for $1$ D0-brane with $2$ D4-branes is computed.
Our result for $2$ D4-branes also agrees with
expectations from a moduli space
analysis~\cite{Kim:2001kp}.\footnote{In~\cite{Kim:2001kp}, the
moduli space is smoothed by turning on an FI term. Our result is
for the case where the FI term vanishes, and there is a small
instanton singularity. A priori, there is no reason for the
counting to agree.} Determining the index and the bulk terms for
arbitrary numbers of D0 and D4-branes remains an outstanding
question. It might be possible to compute the bulk terms
using~\cite{Moore:1998et, Staudacher:2000gx}. To determine the
Coulomb branch, we use a duality between triple
compactifications, and $K3$ surfaces with frozen
singularities~\cite{deBoer:2001px}.

We proceed by applying our binding results to the question of
string/string duality. In the familiar duality between heterotic
on $T^4$ and IIA on $K3$, the IIA string is constructed using a
heterotic $5$-brane wrapping $T^4$. We extend this construction
to the case of the CHL string on $T^4$ versus IIA on a $K3$ with
$8$ frozen $A_1$ singularities~\cite{Schwarz:1995bj}. We show
that the light spectrum for the wrapped $5$-brane agrees with our
expectations for a IIA string on a partially frozen $K3$ surface.
We also provide a new argument for the equivalence of two type I
compactifications: one with a quadruple gauge bundle,\footnote{A
quadruple gauge bundle is a flat bundle on $T^4$ specified by $4$
commuting connections. The bundle is topologically trivial, and
all possible Chern-Simons invariants vanish. However, the bundle
cannot be deformed, while maintaining zero energy, to a trivial
bundle.} and one with a no vector structure (NVS)
compactification. In this approach, the duality becomes completely
geometric.

Our final topic is the construction of domain walls bridging
disconnected string vacua. These vacua each have zero cosmological
constant, and there must exist a field theoretic instanton that
tunnels from one to the other. In Yang-Mills, this instanton
should be BPS carrying fractional charge. What is particularly
nice about this setup is that the vacua are quite simple. Also,
the tunneling involves no change in topology, and so occurs at
finite energy.

When embedded in gravity, there are two modifications: first, we
need to use an instanton/anti-instanton pair to tunnel so the
configuration becomes non-BPS (but stable).\footnote{For a review
of stable non-BPS states, see~\cite{Sen:vx}.} We also expect it to
become time-dependent, with a metric on the wall that looks like
a slice of deSitter space in the thin wall
approximation~\cite{Vilenkin:hy}. Finding CFT/supergravity
solutions for these domain walls would allow us to go beyond
string theory in a fixed background, and perhaps shed light on
questions of cosmology. For prior work on domain walls in the
heterotic string, see~\cite{Singh:2002tf}. While our discussion
is confined primarily to $E_8$ $5$-branes wrapped on $T^3$,
similar phenomena occur for type I D5-branes on $T^4$, and
Euclidean D5-branes on $T^6$ where there are new components in
the string moduli space~\cite{Keurentjes:2001cp,
Morrison:2001ct}.\footnote{A recent discussion of domain walls in
certain $4$-dimensional string compactifications appeared as we
completed this project~\cite{Kachru:2002ns}.} We conclude with a
brief comment on the domain walls we expect in those cases.

\section{Small Instantons in Non-simply-laced Groups}

\subsection{The normalization of instanton charges}

We begin our discussion of instantons by recalling an old theorem
by Bott \cite{Bott:tf}. It states that any continuous mapping of
$S^3$ into a group $G$ can be continuously deformed into a mapping
of $S^3$ into an $SU(2)$ subgroup of $G$. Therefore, as far as
instantons are concerned, we need only study $SU(2)$ subgroups. We
will study the instantons in string theory at loci in the moduli
space where there is enhanced gauge symmetry. Close to such a
locus, any scalars in a vector multiplet act as Higgs fields in
the adjoint representation of $G$. Breaking the gauge symmetry
with adjoint Higgs fields can only result in special subgroups of
$G$, which are called regular subgroups. A regular subgroup has a
root lattice which is a sublattice of the full root lattice of
the group $G$. An adjoint Higgs can be transformed (locally, by a
gauge transformation) into an element of the Cartan subalgebra.
It is then easy to see that the group left unbroken is the one
that commutes with this element, and that it must be regular.
There exists an elegant method due to Dynkin for determining all
regular subgroups of a given group~\cite{Dynkin:um}. However, we
will only be interested in $SU(2)$ subgroups, which can easily be
found by inspection. From these preliminaries, we conclude that
we should study instantons of regular $SU(2)$ subgroups of $G$.

After deformation to an $SU(2)$ subgroup, the instanton charge is
given by, \be \frac{1}{8 \pi ^2 N_{R}} \int~\textrm{Tr}\,(F
\wedge F) = \frac{2 \textrm{Tr}(T^a T^b)}{16 \pi ^2 N_{R}}
\int_{\R^4} \textrm{d}^4x ~{}^* \! F^a_{\mu \nu} F^{b \mu \nu}.
\ee The constant $N_{R}$ is a normalization factor which depends
on which representation, $R$, of the group, $G$, we consider. It
is inserted to normalize the smallest possible instanton charge to
1. For example, if $R$ is the adjoint representation, then
$N_{R}$ is twice the dual Coxeter number. On the right hand side,
we have extracted the generators $T^a$ for the $SU(2)$ subgroup
from the expression. The integral can now be evaluated for an
arbitrary $n$-instanton solution. It is crucial that the charge
is multiplied by $2 \textrm{Tr}(T^a T^b)$ (the factor of 2
ensures that this is integer). One can equate this factor to $k
\delta^{ab}$, with $k$ an integer, known as the embedding or
Dynkin index \cite{Dynkin:um} (see also
\cite{Shifman:ia,Morozov:hy} for related material).

In string theory, where gauge symmetries originate from current
algebra, the integer $k$ is the level of the current algebra in
the current algebra and is called the level. In the operator
product of two currents, $J^a$, the level is the coefficient of
the Schwinger term: \be J^a(z) J^b(w) =
\frac{k\delta^{ab}}{(z-w)^2} + \ldots. \ee
{}For a nice review of
the implications of the Kac-Moody level in string theory,
see~\cite{Dienes:1996du}.

To compute the Dynkin index is not hard: the index for a reducible
$SU(2)$ representation is the sum of the indices of all its
irreducible factors. Let the irreducible representations of
$SU(2)$ be labeled by their dimension $d$, then \be k =
\frac{d(d^2-1)}{6} \ee as can be verified by using the eigenvalues
of $\sigma^3$ in the appropriate irrep.

Let us consider a relevant example. The smallest non-simply-laced
group is $SO(5)\cong Sp(2)$. This group has inequivalent regular
$SU(2)$ subgroups. To find the first one, we decompose $$SO(5)
\rightarrow SO(4) \cong SU(2) \times SU(2)$$ and choose one of the
$SU(2)$ factors. The $\mathbf{4}$, $\mathbf{5}$ and $\mathbf{10}$
(the spin, vector, and adjoint irreps of $SO(5)$) decompose in
the following way, \be \mathbf{4} \rightarrow \mathbf{2 \oplus 1
\oplus 1} \qquad \mathbf{5} \rightarrow \mathbf{2 \oplus 2 \oplus
1} \qquad \mathbf{10} \rightarrow \mathbf{3 \oplus 2 \oplus 2
\oplus 1 \oplus 1 \oplus, 1} \ee and so we find that $k=1,2,6$,
respectively. A charge 1 instanton embedded in this $SU(2)$
subgroup has the smallest charge possible, and indeed one should
set $N_{\bf 4} =1, N_{\bf 5} =2$ and $N_{\bf 10}=6$.

A second $SU(2)$ subgroup can be found by decomposing $$SO(5)
\rightarrow SO(3) \times SO(2).$$ We now embed the instanton in
the $SO(3)$ factor. In this case, the decompositions are given by
\be \mathbf{4} \rightarrow \mathbf{2 \oplus 2} \qquad \mathbf{5}
\rightarrow \mathbf{3 \oplus 1 \oplus 1} \qquad \mathbf{10}
\rightarrow \mathbf{3 \oplus 3 \oplus 3 \oplus 1}. \ee This gives
$k=2,4,12$, respectively. The instanton charge in this case is
twice as large as in the previous example. It is not hard to see
that if $SO(5)$ is broken with only adjoint Higgs fields, only the
second breaking is possible. Therefore, any breaking of $SO(5)$
by adjoint Higgs fields can preserve, at most, instanton
configurations with even charge.

The previous discussion can be rephrased in a slightly more
abstract way in terms of properties of the group lattice of the
non-simply-laced group. This will allow us to generalize to the
case of a semi-simple gauge group with factors at different
levels. In either case, by definition, there are roots of
different lengths. We can focus on the case where there are two
different lengths, with the generalization clear. By standard
group theory, we can associate an $SU(2)$ subgroup to every root,
and these are the groups in which we will embed elementary
instantons.

Any two $SU(2)$ subgroups corresponding to roots of different
lengths clearly cannot be conjugate. Hence their instanton
solutions are in general also inequivalent, no matter where we
are in the moduli space. The various representations of $G$ are
generated by vectors in the weight lattice. The lattice dual to
the weight lattice, known as the coroot lattice\footnote{The
coroot lattice is the root lattice of the dual group,
$\widehat{G}$. For non-simply-laced groups, the group and the dual
group differ.}, determines the global structure of any subgroup
that we might choose to study. The periodicity of any generator
for an $su(2)$ subalgebra can be determined from the coroot
lattice.

A long root corresponds to a short coroot, and vice versa.
Suppose one has a root with length $p$, and another root with
length $q$. The same instanton solution, embedded in either of
the $SU(2)$ subgroups associated to these roots, will result in
inequivalent solutions whose charges have a ratio $q/p$. The
solution of smallest charge occurs in the subgroup associated to
the largest root (and therefore the smallest coroot). In the
above $SO(5)$ example, it is not hard to verify that the two
different solutions correspond to decompositions which use either
a long or a short root of the $SO(5)$ algebra, respectively.

Note that the multiplicative normalization of the instanton
charge is proportional to the level, but inversely proportional
to the lengths of the roots of the various subgroups. We can now
embark on a study of instantons in string theories where
non-simply-laced groups, and groups at different levels appear.

\subsection{Instantons and orientifolds}

Much of what we have just described can be understood more
intuitively using orientifolds.  We use the conventions
of~\cite{deBoer:2001px}\ where an \Op{p}-plane together with $N$
D$p$-branes supports an $Sp(N)$ gauge group. An \Om{p}-plane with
$N$ D$p$-branes supports an $SO(2N)$ group, while \Omt{p}
supports an $SO(2N+1)$ gauge group.

Consider \Op{4} with $N$ pairs of coincident D4-branes. The
orientifold plane supports an $Sp(N)$ gauge group, which is
non-simply-laced. Instantons in this gauge theory can be realized
by D0-branes which are stuck to the orientifold plane. It is
natural for us to ask about the dynamics of D0-brane probes of
the orientifold plane. The quantum mechanical gauge theory on $k$
D0-brane probes of $O4^+$ has $8$ real supercharges. It has gauge
group $O(k)$ with a hypermultiplet transforming as a rank $2$
symmetric tensor. The global symmetry of the theory is $Sp(N)$,
and the D0-D4 strings give a half-hypermultiplet transforming in
the bifundamental of $O(k)\times Sp(N)$. The Higgs branch of the
theory is the moduli space of $k$ $Sp(N)$ instantons. The case
$k=1$ corresponds to a $\Z_2$ gauge group. A single D0-brane can
be stuck at the location of the orientifold plane, and cannot
move into the bulk. Therefore, there is no Coulomb branch. This is
true for any choice of $N$, including $N=0$. In this case, the
D0-branes correspond to ``Sp(0)'' instantons.

\begin{figure}[h]
\begin{center}
\includegraphics[width=14cm]{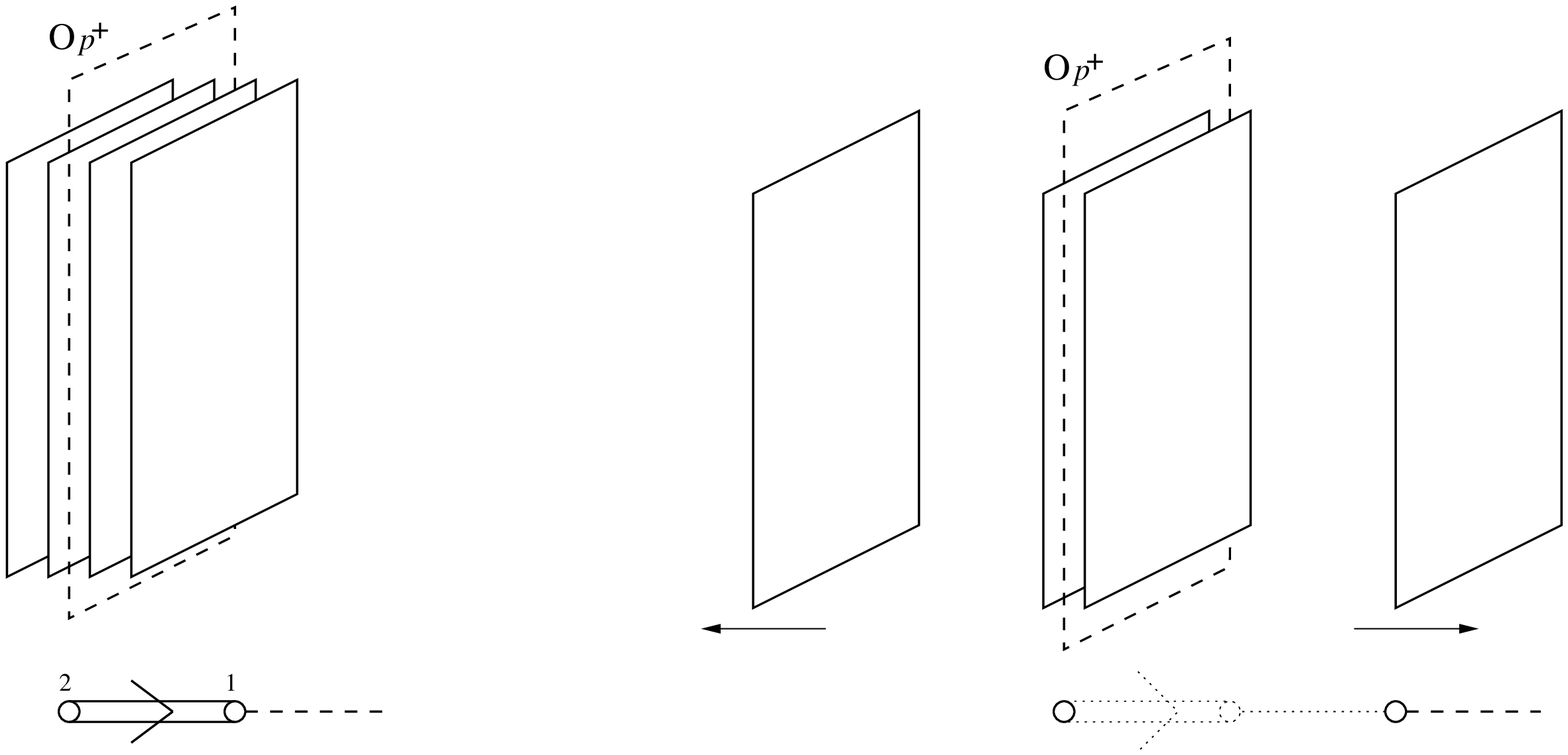}
\caption{$Sp(N)$: the long root should be identified with the
orientifold plane.}
\end{center}
\end{figure}
That a single D0-brane can localize at an \Op{4}-plane can also be
understood by interpreting the D0-branes as small instantons in
the D4 gauge theory. Since $Sp(N)$ is non-simply-laced, it has
roots of different length. We can move pairs of D4-branes away
from the orientifold plane, but the group at the \Op{4} remains
symplectic. A comparison with the Dynkin diagram of $Sp(N)$
suggests that we should associate the long root of the Dynkin
diagram with the \Op{4}-plane. For a similar observation,
see~\cite{Hanany:2001iy}. What is important for us is the
periodicity of the group, which is determined by the coroots. The
long roots of $Sp(N)$ give rise to short coroots. Instantons with
the smallest possible charge have to be embedded in the $SU(2)$
associated with the short coroots. To these instantons, we assign
charge one. By contrast, it is natural to assign charge two to
instantons living in the $U(N)$ theory on bulk D4-branes
separated from the \Op{4}-plane. If there were no \Op{p}-plane,
these instantons would have the smallest charge possible, and it
would then be natural to assign them charge one.

\subsection{Can a D0-brane stick to an \Op{4}-plane?}
\label{Boundstatecomp}
\subsubsection{Symmetries and supercharges}

Let us take $k=2$ with $N=0$. This particular case will play a
role in later discussion. The D0-brane gauge theory now has a
Coulomb branch, and so the probe can move in the bulk. Although
there is no gauge group localized at the orientifold plane, the
D0-brane might still bind to the orientifold plane. To address
this question, we use $L^2$ index theory.

In addition to an $O(2)$ vector multiplet, there is one
hypermultiplet transforming as a second rank symmetric tensor;
see, for example,~\cite{Hori:1999me}. We can factor out the trace,
which gives a decoupled hypermultiplet. This hypermultiplet
parametrizes the position of the D0-brane along the \Op{4}-plane.
What remains are $2$ hypermultiplets with charge $2$ under the
$SO(2)$ subgroup of $O(2)$. We anticipate however that in the
computations below, the charge will not make any difference. We
will therefore denote it by $e_0$, and demonstrate explicitly
that our end result does not depend on this variable. The
symmetry group for $2$ hypermultiplets is $Sp(2)_L\times
Sp(1)_R$. We realize the R-symmetry via the $Sp(1)_R$ action. The
gauge symmetry commutes with the R-symmetry, and so must sit in
$Sp(2)_L$. Note that a single hypermultiplet is not possible in
this case because $O(2)$, unlike $SO(2)$, cannot be faithfully
embedded in $Sp(1)_L$. However, we can embed $O(2)$ into
$Sp(2)_L$. The full symmetry group of the theory is the
combination of the dimensionally reduced Lorentz group and the
R-symmetry, $Spin(5)\times Sp(1)_R$. Our conventions follow those
of~\cite{Sethi:2000ba}.

The vector multiplet contains $5$ scalars, $X^\m$, which are in
the $(\bf{5}, \bf{1})$ of the symmetry group. These scalars are
inverted by the $O(2)$ group element, \be \label{go} g_{\rm o} =
\begin{pmatrix}{1 & 0 \cr 0 & -1 } \end{pmatrix}, \ee but
uncharged under the $SO(2)$ subgroup. Let $P^\mu$ be the
associated canonical momenta obeying, \be\label{bosquant}
\com{X^\mu}{P^\nu} = i\delta^{\mu\nu}. \ee The superpartners of
these bosons are eight real fermions, $\lambda_a$, where
$a=1,\ldots,8$ transforming in the $( \bf{4}, \bf{2})$
representation. These fermions obey the usual quantization
relation, \be\label{fermquant}
 \acom{\lambda_{a}}{\lambda_{b}} = \delta_{ab}.
\ee We also need hermitian real gamma matrices, $\gamma^\m$, which
obey \be\label{gmat} \acom{\gamma^\mu}{\gamma^\nu} = 2
\delta^{\mu\nu}. \ee To complete the vector multiplet, we
introduce an auxiliary field, $D$, which transforms as $({\bf
1},{\bf 3})$ under the symmetry group. Supersymmetry requires
that it be an imaginary quaternion, independent of $X^\mu$.
The vector multiplet supercharge is given by: \be
\label{vectorsuper} Q^v = \ps \lambda + \frac{e_0}{2} D \lambda.
\ee

A hypermultiplet contains four real scalars which we can package
into a quaternion $q$ with components $q^i$ where $i=1,2,3,4$.
This field transforms as $({\bf 1}, {\bf 2})$ under the symmetry
group. We again introduce canonical momenta $p_i$ satisfying the
usual commutation relations.

We have $2$ hypermultiplets, $q$ and $\tq$ which have charge $e_0$
and charge $-e_0$, respectively, under the $U(1)$ subgroup of
$O(2)$. We generate gauge transformations on the bosons (packaged
into two quaternions)  using left multiplication by, $$
\pmatrix{q \cr \tq} \rightarrow I \pmatrix{0 & -\bf{1} \cr \bf{1}
& 0} \pmatrix{q\cr \tq}.$$ This action is realized by the
operator, \be G_B =  \hlf \left(\tq \bar{p} + p \bar{\tq} - q
\bar{\tp} - \tp \bar{q}\right), \ee which is real as a
quaternion, and therefore hermitian with respect to the
components of the quaternion.

To go from $U(1)$ to $O(2)$, we will also need to gauge the $\Z_2$
symmetry corresponding (in a complex basis) to charge conjugation.
We will return to this point momentarily. The superpartner to
$(q, \tq)$ is a real fermion $(\f_a, \tf_a)$ with $a=1,\ldots, 8$
satisfying, \be\label{secondquant} \acom{\f_{a}}{\f_{b}} =
\delta_{a b}, \quad \acom{\tf_{a}}{\tf_{b}} = \delta_{a b}, \quad
\acom{\f_{a}}{\tf_{b}} = 0 \ee and transforming in the $({\bf 4},
{\bf 1})$ representation. Converting the $p_i$ to quaternions,
with the aid of the $s^j$ operators given in Appendix A, the free
hypermultiplet charge takes the form \be\label{freehyper} Q^{h_f}
= p \f + \tp \tf. \ee This free charge obeys the algebra, $$
\acom{Q^{h_f}_a}{Q^{h_f}_b} = \delta_{ab}(|p|^2 + |\tp|^2).$$
Invariance of \C{freehyper}\ under the $U(1)$ gauge symmetry
requires that \be\label{gaugefermion} G_F = i\, (\tf_a \f_a) \ee
generate gauge transformations on $\f, \tf$. The total generator
of the $U(1)$ subgroup of the gauge symmetry is then,
\be\label{gauge} G = G_B + G_F = \hlf \left( \tq \bar{p} + p
\bar{\tq} - q \bar{\tp} - \tp \bar{q} \right) + i \tf \f. \ee The
full hypermultiplet supercharge $Q^h$ also includes couplings to
the vector multiplet, \be\label{hypercharge} Q^h = p \f + \tp \tf
+  \frac{e_0}{2}\xs (q \tf - \tq \f). \ee Note that the order of
multiplication matters because $p, \xs ,q$ are matrix-valued
fields. The form of the interaction term in \C{hypercharge}\ is
fixed up to an overall constant by symmetry. The charge obeys the
algebra: \be\label{hyperalg} \acom{Q^h_{a}}{Q^h_{b}} =
\delta_{ab} \left\{ |p|^2 +|\tp|^2  + \frac{e_0^2}{4}|X|^2
(|q|^2+|\tq|^2) -  i e_0 \, (\f \! \xs \tf) \right\} +
\frac{e_0}{2}(\xs)_{ab} G. \ee Closure of the supersymmetry
algebra is, as usual, only up to gauge transformations.

The full supercharge $Q$ is the given by, \be\label{fullcharge} Q
= Q^v + Q^h, \ee where we define the $D$-term in the following
way: \be \label{Dterm} D = \frac{1}{2}(q \bar{\tq}- \tq \bar{q}).
\ee The full charge obeys the algebra: \be\label{completesusy}
\acom{Q_{a}}{Q_{b}} = 2 \delta_{ab} (H_B + H_F + V) +
\frac{e_0}{2}\xs_{ab} G \ee with \bea
 & H_B & =  \frac{1}{2} \left( P^2 + |p|^2 + |\tp|^2 \right), \\
 & H_F & =  i \frac{e_0}{2} \left( \lambda \tq \f - \lambda q \tf - \f \! \xs
\tf \right), \\
 & V & =   \frac{e_0^2}{8} \left( |X|^2 \{|q|^2+|\tq|^2 \} + |D|^2
 \right).
\eea

Lastly, we need to check that the supercharges are $O(2)$ rather
than $SO(2)$ invariant. Any element of $Sp(2)_L$ must preserve
the norm, $$ N(q, \tq) = q \bar{q} + \tq \bar{(\tq)}. $$ Consider
the element $$ \pmatrix{\bf{1} & 0  \cr 0 & -\bf{1}} $$ which
squares to one. This group element is trivially identified with
the generator, $g_o$, given in \C{go}. Under its action, \be
\label{Otwo} (q, \f) \leftrightarrow (q,\f) \quad (\tq, \tf)
\leftrightarrow (-\tq, -\tf), \quad X \rightarrow -X, \quad
\lambda \rightarrow -\lambda. \ee It is easy to see that $D
\rightarrow -D$ so the vector multiplet charge is invariant. It
is also easy to see that the hypermultiplet charge is invariant
so this is a symmetry of the theory which we can gauge.

Finally we note that the centralizer of $O(2)$ inside $Sp(2)$ are
matrices of the form $$ \pmatrix{ S & 0 \cr 0 & S } $$ with $S$
an element of $Sp(1)$. This diagonal $Sp(1)_f$ flavour symmetry,
acting from the left, can be combined with the $R$ symmetry
$Sp(1)_R$ that acts from the right. Together they give an $SO(4)$
symmetry under which the $4$ components $(q_i, \tq_i)$ of each
quaternion transform as a vector.

This theory has two branches. The Coulomb branch is parametrized
by the $X^\m$, and is $\R^5/\Z_2$. At a generic point, the
discrete $\Z_2$ symmetry is broken, and the gauge group $SO(2)$.
 The Higgs branch is obtained by setting $\tq =
0, \, X=0$ (or equivalently by an $SO(2)$ gauge transformation, $q
= X=0$), and quotienting by the residual $\Z_2$ gauge symmetry,
giving $\R^4/\Z_2$. On this branch it is the $SO(2)$ symmetry
which is broken to $\Z_2$. This leaves a residual $O(1) \times
O(1) = \Z_2 \times \Z_2$ gauge symmetry. No FI term is possible
in this theory because any allowed $D$-term must be odd under
$g_o$. A constant $D$-term is therefore ruled out.

The Coulomb branch parametrizes motion of the D0-brane away from
the orientifold plane. The Higgs branch occurs because at the
\Op{4}-plane, half-integer charged D0-branes are possible. The
Higgs branch corresponds to the splitting of a bulk \D{0}-brane
into two such half \D{0}-branes. The expectation value of the
scalars in the hypermultiplet corresponds to the separation of
the two fractionally charged constituents along the \Op{4}-plane.
There is also a decoupled hypermultiplet that describes the
center of mass motion along the \Op{4}-plane. Note that the
residual gauge symmetry on the Higgs branch is exactly what we
expect for two D0-branes of charge one-half.

\subsubsection{The bulk term contribution}

To compute the $L^2$ index, we need to evaluate the
low-temperature limit of the twisted partition function \be {\rm
Ind}=  \lim_{\beta\rightarrow\infty} {\rm Tr} (-1)^F e^{-\beta H}.
\ee This is a topic that has been analyzed in some detail, and we
will use and extend the methods developed in~\cite{Sethi:1996kj,
Sethi:1997pa, Yi:1997eg}.

We begin by computing the bulk term contribution which is the
high temperature limit of the twisted partition function: \be
I(0) = \lim_{\beta\rightarrow 0} {\rm Tr} (-1)^F e^{-\beta H}.
\ee The $\Z_2$ charge measuring fermion number is given by \be
(-1)^F = 2^{12}\prod_{a=1}^{8} \lambda_a \prod_{a=1}^{8}
\f_a\prod_{a=1}^{8} \tf_a. \ee We need to approximate the heat
kernel $e^{-\beta H}$, but fortunately, the simplest
approximation will suffice: \be e^{-\beta H}(X,(q,\tq)
;X',(q',\tq') ) = {1\over (2\pi\beta)^{13/2}} \, e^{ \,- {1\over
2 \beta} \left(|X-X'|^2 + |q-q'|^2 +| \tq - \tq' |^2 \right)}
e^{-\beta V} e^{-\beta H_F}(1+ O(\beta)). \ee We have lumped all
the bosonic potential terms into $V$. We also need to be sure
that we compute the trace on gauge invariant states so we insert
a projection operator into the trace, \be\label{bulk} I(0) =
\lim_{\beta\rightarrow 0} \, \int_{-\pi}^{\pi} {d\theta\over 2
\pi}\, {\rm Tr} (-1)^F {1\over 2} (1+ \Pi (g_o)) e^{i e_0 \theta
G_F} e^{-\beta H}(X,(q,\tq) ; X, g_{e_0\theta} (q,\tq)), \ee where
$$g_{e_0\theta} (q, \tq) = (\cos e_0 \theta \ q - \sin e_0\theta
\ \tq, \, \sin e_0\theta \ q + \cos e_0\theta \ \tq)$$ The
projection onto $O(2)$ invariant states is performed by the
insertion of, ${1\over 2} (1+ \Pi (g_o))$, where $\Pi(g_o)$
implements the action of $g_o$ on all the fields.

There are two contributions to the bulk term \C{bulk}, one with
$\Pi(g_o)$ inserted, and one without. Let us first deal with the
case where $\Pi(g_o)$ is inserted. Note that $\Pi(g_o)$ sends
$X\rightarrow -X$, and so leaves us with an approximate heat
kernel, $$ e^{-\beta H} \sim \left( {1\over \beta}\right)^{13/2}
e^{-{2\over \beta}|X|^2} e^{\,- {1\over 2 \beta}
\left(|q-g(\theta)(q)|^2 +| \tq - g(\theta)(\tq) |^2 \right)} e^{-
{{\beta e_0^2}\over 8} \left( |X|^2 (|q|^2 + |\tq |^2 ) + |D|^2
\right)} e^{\left(i e_0\theta G_F -\beta H_F\right)}. $$ We need
to saturate the trace with fermions from the propagator.
Otherwise, the insertion of $(-1)^F$ kills the trace. From the
perspective of the Euclidean path-integral, inserting $(-1)^F$
means that the fermions have periodic boundary conditions in the
time direction. Consequently, there are fermion zero-modes. Since
$\Pi(g_o)$ also sends $$\lambda\rightarrow -\lambda,\quad \tf
\leftrightarrow -\tf,$$ there are no $\lambda$ or $\tf$ zero
modes, but there are $8$ $\f$ zero modes. We now need to count
powers of $\beta$. To prevent the integral from vanishing, we
need to rescale $x \rightarrow {x\over \sqrt{\beta}}$ which
introduces a factor of $\beta^{5/2}.$ We also rescale
$\theta\rightarrow {\theta\over\sqrt{\beta}}$ which introduces
$\beta^{1/2}$. Finally, we need to rescale the combination $\tq
\rightarrow {\tq \over\sqrt{\beta}}$ which gives $\beta^{4/2}$.
Schematically, what remains takes the form $$ \left({1\over
\beta}\right)^{3/2} e^{ -\beta^{1/2}\theta \F \F - \beta^{3/2} X
\F \F - \beta q \F\lambda + \beta^{3/2}\tq \F\lambda} $$ where
$\F$ denotes both $\f, \tf$.  To saturate the $\f$ zero modes
requires at least $4$ insertions of a $\F\F$ term. This brings
down a minimum of $\beta^2$, which kills this contribution.

We therefore need only consider the bulk term without any
$\Pi(g_o)$ insertion. The bulk term for this $U(1)$ gauge theory
only differs by a factor of $2$ from the computation for $O(2)$,
$$ I(0)_{U(1)} = 2 I(0)_{O(2)}.$$ In the limit $\beta\rightarrow
0$, we can localize $g(\theta)$ around the identity, and make the
replacement \be e^{\,- {1\over 2 \beta} \left(|q-g(\theta)(q)|^2
+| \tq - g(\theta)(\tq) |^2 \right)} \quad \rightarrow \quad
e^{\,- {e_0^2 \theta^2 \over 2\beta} \left( |q|^2 + | \tq |^2
\right)}.\ee At this point, we examine the charge dependence of
the index computation. The relevant part of the heat kernel has
the form, $$ \int_{-\pi}^{\pi} \, {d\theta\over 2 \pi} \,e^{-|Q -
e^{ie_0\theta} Q|^2}\times \ldots, $$ where $Q$ represents the
matter fields. There is a non-zero contribution from a
neighborhood of each solution of $$ e^{ie_0\theta}=1.$$ There is
only one special case which occurs when the boundary points,
$\theta=\pm \pi$, are solutions. These points together give the
same contribution as an interior point, $\theta \in (-\pi, \pi)$.
The total contribution is then $|e_0|$ times larger than the case
of matter with charge one.

With these comments in mind, we note that the rescaled gauge
parameter, $\tilde{\theta}= \theta/\beta$, then effectively
behaves like the $A_0$ component of the gauge-field in Euclidean
space. Upon changing variables from $\theta$ to $\tilde{\theta}$,
we get an additional factor of $\beta$ in the measure of our
integral. From the proceeding discussion, we know that we can
account for the charge dependence by simply multiplying the
$\theta=0$ contribution by $e_0$.

In the $\beta \rightarrow 0$ limit, the range of $\tilde{\theta}$
diverges, and we can really treat $\tilde{\theta}$ on equal
footing with $X^{\mu}$. If we express the fermion bilinear
appearing in $\beta H_F - i e_0\theta G_F$ in the form $\Psi M
\Psi$, where $\Psi$ collectively denotes $\lambda, \f,\tf$ then
$M$ is a matrix linear in the bosons. It takes the form, $$
\frac{i e_0 \beta}{4} \pmatrix{\lambda & \f & \tf} \pmatrix{0 &
\tq & -q \cr -\tq^T & 0 & -\xs - i\tilde{\theta} \cr q^T & \xs
+i\tilde{\theta} & 0 } \pmatrix{\lambda \cr \f \cr \tf}.$$ Our
task is to determine the Pfaffian. It is $SO(6)$ invariant with
$(X^{\mu},\tilde{\theta})$ forming a vector $Y$ under the
$SO(6)$. The Pfaffian can therefore only depend on $|Y|=\sqrt{X^2
+ \tilde{\theta}^2}$. To find the scaling behaviour with respect
to various contributions, we use the following trick.

Multiply the first and second (quaternionic) row  with a positive
real number, $\rho$, and the third row with $\rho^{-1}$. Then
multiply the first and second column with $\rho$, and the third
with $\rho^{-1}$. The net effect of these manipulations is $$ Y
\rightarrow Y, \quad  q \rightarrow q,\quad \tq \rightarrow
\rho^2 \tq, \qquad {\rm Pf}(M) \rightarrow \rho^8 {\rm Pf}(M). $$
{}From this we deduce that the Pfaffian must contain 4 powers of
$\tq$. Repeating these manipulations with different combinations
of rows and columns, one also finds that the Pfaffian contains $4$
powers of $q$, and $4$ of $|Y|$.

To find the Pfaffian, we make use of symmetry. First we use the
$SO(6)$ symmetry to rotate to coordinates where $X^{\mu} = 0$ and
$\tilde{\theta}= |Y|$. Next we make use of the $Sp(1)_f \times
Sp(1)_R = SO(4)$ which acts on the quaternions $q$ and $\tq$.
First act with the orthogonal matrix: $$ |q|^{-1} \pmatrix{ q_1 &
q_2 & q_3 & q_4 \cr
                     -q_2 & q_1 &-q_4 & q_3 \cr
                     -q_3 & q_4 & q_1 &-q_2 \cr
                     -q_4 &-q_3 & q_2 & q_1 }
\pmatrix{ q_1 & \tq_1 \cr q_2 & \tq_2 \cr q_3 & \tq_3 \cr q_4 &
\tq_4} = |q|^{-1} \pmatrix{ |q|^2   & q_i \tq_i \cr 0 & -D_1 \cr
0 & -D_2 \cr 0 & -D_3 }. $$ Note that this matrix effectively
implements right multiplication by $\bar{q}/|q|$ which is a unit
quaternion. The $D_i$ are the components of the $D$-term, which
is the imaginary part of $-\tq \bar{q}$; see \C{Dterm}. Now use
an $SO(3)$ rotation which leaves the first row invariant to set
$$ \tq \rightarrow |q|^{-1} (q_i \tq_i s^1 + |D| s^2).$$ The
matrix $M$ now takes a very simple form. By row and column
manipulations, we can make it block diagonal, with two $6 \times
6$ blocks of the form $$ M' = \frac{i e_0\beta}{4} \pmatrix{ 0 &
0 & \frac{q_i \tq_i}{|q|} & - \frac{|D|}{|q|} & -|q| & 0  \cr 0 &
0 &  \frac{|D|}{|q|}  & \frac{q_i \tq_i}{|q|} &  0  &-|q| \cr
-\frac{q_i \tq_i}{|q|} & -\frac{|D|}{|q|} & 0 & 0 & -|Y| & 0  \cr
 \frac{|D|}{|q|} & -\frac{q_i \tq_i}{|q|} & 0 & 0 &  0  &-|Y| \cr
 |q| & 0 & |Y| & 0 & 0 & 0 \cr
 0   & |q| & 0 & |Y| & 0 & 0}, $$
and two 6 x 6 blocks having the same form except that $|D|$ is
replaced by $-|D|$.

We can now compute the Pfaffian, \be {\rm Pf}(M) = \left( {\rm
Pf}(M') \right)^4 = \frac{(e_0\beta)^{12}}{2^{20}}  |Y|^4 |D|^4.
\ee This is clearly invariant under the required symmetries, and
has the predicted scaling behaviour.

Tracing over the fermions leaves us with the Pfaffian multiplied
by a factor of ${\rm Tr}(I)=2^{12}$ from the identity operator
acting on the fermion Hilbert space. Having dealt with the
fermions, the bulk integral becomes \be
\frac{e_0^{13}\beta^{13/2}}{2^{10} (2 \pi)^{15/2}}\int \dif^6 Y \
\dif^4 q \ \dif^4 \tq \ |Y|^4|D|^4 e^{ -\frac{e_0^2
\beta}{8}(|Y|^2(|q|^2+ |\tq|^2) + |D|^2)}.\ee Rescaling $X,
\tilde{\theta}, q, \tq$ with $(e_0^2\beta/4)^{1/4}$, the $\beta$
and $e_0$ dependences drops out of the integral. The integral
over $Y$ can be converted to six-dimensional polar coordinates.
The result is (noting that the volume of a $5$-sphere is $\pi^3$)
$$ \frac{3\cdot 2^7}{(2 \pi)^{9/2}} \int \dif^4 q \ \dif^4 \tq \
\frac{|D|^4}{\left(|q|^2 + |\tq|^2\right)^5} \ e^{-\frac{1}{2}
|D|^2}. $$

To proceed, we note that $$ |D|^2 = \hlf \sum_{ij}(q_i \tq_j -
q_j \tq_i)^2 = |q|^2 |\tq|^2 \sin^2 \alpha, $$ where $\alpha$ is
the angle between $q$ and $\tq$ as 4-vectors. We now convert the
integrals over $q$ and $\tq$ both to 4-dimensional polar
coordinates. In such coordinates, the measure becomes $$ \dif^4 x
= r^3 \ \dif r \ \dif \Omega_3 = r^3 \sin^2 \alpha' \ \dif r \
\dif \Omega_2 \ \dif \alpha'. $$ Here $\dif \Omega_n$ denotes the
measure for an $n$-sphere. The coordinate $\alpha'$ ranges from
$0$ to $\pi$. If we think of the $\tq$ integral as being inside
the $q$ integral, we can choose polar coordinates for the $\tq$
integral such that the direction of the ``south-pole'', the
points with $\alpha'= 0$, corresponds to the direction of $q$.
Then we can simply identify the angle $\alpha$ between $q$ and
$\tq$ with $\alpha'$.
\begin{figure}[h]
\begin{center}
\includegraphics[width=6cm]{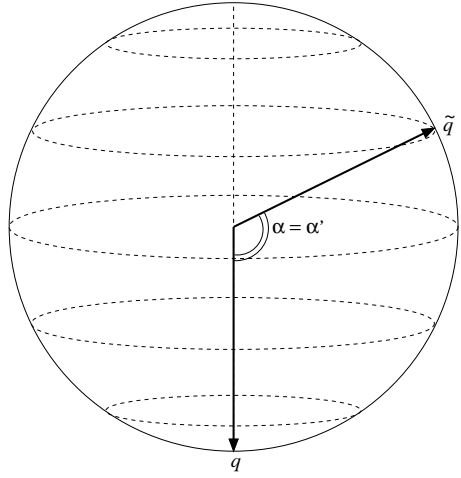}
\caption{The geometry for the integrals. One dimension is
suppressed: the latitude circles are actually $2$-spheres which
scale with $\sin^2 \alpha$ rather than $\sin \alpha$.}
\end{center}
\end{figure}

Integrating over the 3- and 2-spheres in the $q$ (resp. $\tq$)
integral gives factors of $2 \pi^2$ (resp. $4 \pi$). We also make
the substitutions $r=|q|^2$, $\tilde{r}=|\tq|^2$, which result in
the integral $$\frac{3 \cdot 2^7}{(2 \pi)^{3/2}} \int_0^{\infty}
\dif|q| \int_0^{\infty} \dif |\tq| \int_0^{\pi} \dif \alpha \
\frac{|q|^7|\tq|^7 \sin^6 \alpha}{\left(|q|^2 + |\tq|^2\right)^5}
\ e^{-\hlf |q|^2|\tq|^2 \sin^2 \alpha}= $$ $$\frac{3\cdot 2^5}{(2
\pi)^{3/2}} \int_0^{\infty} \dif r \int_{0}^{\infty} \dif
\tilde{r} \int_0^{\pi} \dif \alpha \ \frac{r^3 \tilde{r}^3 \sin^6
\alpha}{\left(r + \tilde{r}\right)^5} \ e^{-\hlf r\tilde{r}
\sin^2 \alpha}. $$ Substituting $x=\tilde{r}/r$, the integrals
over $r$ and $\alpha$ become simple: $$ \frac{3\cdot 2^5}{(2
\pi)^{3/2}}  \int_0^{\infty} \dif r \int_{0}^{\infty} \dif x
\int_0^{\pi} \dif \alpha \ \frac{r^2 x^3 \sin^6 \alpha}{\left(1 +
x \right)^5} \ e^{- \hlf r^2 x \sin^2 \alpha}=$$ $$
\frac{3\cdot2^3}{\pi} \int_{0}^{\infty} \dif x \int_0^{\pi} \dif
\alpha \ \frac{x^{3/2} \sin^3 \alpha}{\left(1 + x \right)^5}=
\frac{2^5}{\pi} \int_{0}^{\infty} \dif x \ \frac{x^{3/2}}{\left(1
+ x \right)^5}$$ Upon substituting $x = \tan^2 \phi$, the final
integral becomes solvable: $$ \frac{2^6}{\pi} \int_{0}^{\pi/2}
\dif \phi \ \cos^4 \phi \ \sin^4 \phi = \frac{1}{\pi}
\int_{0}^{\pi/2} \dif \phi \left(\frac{3}{2} - 2 \cos 4 \phi +
\frac{\cos 8 \phi}{2} \right). $$ Therefore the bulk term for the
$O(2)$ theory is, \be \label{totalbulk} I(0)_{O(2)} = \frac{3}{4}.
\ee

\subsubsection{The defect term contribution}

The defect term comes from two sources in this problem. Both are
boundary terms, and it is easy to see that they are independent
of the charge $e_0$. The first comes from the boundary of the
Coulomb branch. The contribution from this source is given by the
defect contribution of a free particle moving on the moduli space
$\R^5/\Z_2$. This is precisely the same computation as appears in
the study of $SU(2)$ quantum mechanical gauge theory with $8$
supercharges. The contribution is~\cite{Yi:1997eg, Sethi:1997pa},
\be\label{defectone} I_D^{(1)} = - {1\over 4}. \ee

The second contribution comes from the boundary of the Higgs
branch. Without a detailed justification (along the lines given
in~\cite{Sethi:1997pa}), we can compute this contribution by
studying the defect for a free particle moving on $\R^4/\Z_2$. We
trace over the difference between even and odd wavefunctions,
which depend on the $4$ light $q^i$. There is a degeneracy of $8$
coming from quantizing the $8$ light fermions. Together these
factors give, \be\label{defecttwo} I_D^{(2)} =  - 8
\lim_{\beta\rightarrow 0} \int \, { d^4q \over (2\pi \beta)^2} \,
e^{-{|q+q|^2 \over 2 \beta}} = - {1\over 2}. \ee

Collecting the results from \C{totalbulk},\C{defectone}, and
\C{defecttwo}, we find two results. For the case of $U(1)$ with
two charged hypermultiplets, we find that the $L^2$ index is \be
{\rm Ind}_{U(1)} = 2I(0) + I_D^{(2)} = \frac{3}{2} -\frac{1}{2} =
1. \ee This corresponds to the case of a D0-brane probing two
D4-branes. Note that in this case there is no contribution from
the Coulomb branch which is $\R^5$ rather than $\R^5/\Z_2$. The
answer is the same as the case of a D0-brane probing a single
D4-brane obtained in~\cite{Sethi:1996kj}.

For the case of $O(2)$ with a symmetric hypermultiplet, we find
that the index is \be {\rm Ind}_{O(2)} = I(0) + I_D^{(1)} +
I_D^{(2)} = \frac{3}{4} - \frac{1}{4} -\frac{1}{2} = 0. \ee This
corresponds to the case of a D0-brane probing \Op{4}. We note
that the invariance theorem of~\cite{Sethi:2000zf}\ implies that
the index, in both cases, actually counts the total number of
bound states. So in this case, there is no bound state at all. It
is interesting that there is no bound state despite the existence
of a Higgs branch. We shall require these results later in
checking various string-string dualities.

\subsection{Comments on \Omt{p}}

The other case of a non-simply-laced gauge group is associated
with an \Omt{4}-plane. However, the physics of D0-brane near an
\Omt{4}-plane is entirely different from the \Op{4} case. The
gauge theory on a D0-brane near an \Omt{4}-plane with $N$ pairs of
coincident D4-branes has $8$ supercharges. It consists of an
$Sp(1)$ vector multiplet, a decoupled hypermultiplet in the
antisymmetric representation, and a half-hypermultiplet in the
bifundamental representation of the symmetry group, $Sp(1)\times
O(N)$.

The most interesting situation is when there is one pair of \D{4}
branes coincident with the \Omt{4}-plane, so that the spacetime
gauge group is $O(3)$. Now consider the situation where there is
a bulk D0-brane in the vicinity of the orientifold plane. We
claim that the moduli space for the \D{0}-brane does not have a
Higgs branch!

One can see why this is so in various ways. The most direct way is
by explicitly solving for the D0-brane moduli space, and checking
that there is no non-trivial Higgs branch. As noted
in~\cite{Witten:1995gx}, the construction of the moduli space for
$k$  \D{(p-4)} pairs near an \Om{p} with $N$ D$p$-branes parallels
the ADHM construction for orthogonal groups described, for
example, in~\cite{Christ:jy}. The authors of~\cite{Christ:jy}\
point out that for $O(3)$, their construction is problematic. In
this case, it appears to depend on the wrong number of
parameters. The counting suggests that $k/2$ rather than $k$
should be identified with the topological charge. This assertion
is true, but we will nevertheless identify $k$ with the \D{0}
brane charge for the following reasons.

Instantons are characterized by the homotopy class of maps from
the $S^3$ which bounds Euclidean $\R^4$ to the gauge group $G$.
All simple Lie groups $G$ have $\pi_3(G)= \Z$. We can always view
$\pi_3$ as generated by a map from $S^3$ into an $SU(2)$ subgroup
of $G$. Consider $O(3)$, or rather its connected component,
$SO(3)$. Since $SO(3)= SU(2)/\Z_2$, and $SU(2) \equiv S^3$ as a
group manifold, the instantons come from  maps \be S^3 \rightarrow
SO(3) \equiv S^3/\Z_2 . \ee The natural maps from $S^3$ to
$SO(3)$ are those that wind the group $S^3$ an even number of
times around $S^3/\Z_2$ (thereby essentially treating $S^3$ as two
copies of $S^3/\Z_2$). In this sense, the map with winding number
2 generates $\pi_3(SO(3))$. Of course, we could simply divide
everything by 2, thereby mapping the even numbers to the
integers, but there is a good reason not to do so.

We can view the $SO(3)$ gauge group on \Omt{4} with 1 pair of
D4-branes as a subgroup of the case with $N>1$ D4-branes and
gauge group $SO(2N+1)$. Again $\pi_3(SO(2N+1))= \Z$, but this
group is not generated by the $\pi_3$ of the $SO(3)$ subgroup. The
$SU(2)$ subgroup whose homotopies generate the fundamental group
of $SO(2N+1)$ is a subgroup of $SO(4)= \left\{SU(2) \times SU(2)
\right\} / \Z_2$. Rather, the instanton solution of $SO(3)$
generates the subgroup of $\pi_3$ associated with even windings.
So even if we consider just an $SO(3)$ group, we should normalize
the elementary instanton in $SO(3)$ to have charge $2$ because we
can continuously deform this theory to one with $SO(2N+1)$ gauge
symmetry. In brane language, we can smoothly bring D4-branes in
from infinity.
 \begin{figure}[h]
\begin{center}
\includegraphics[width=14cm]{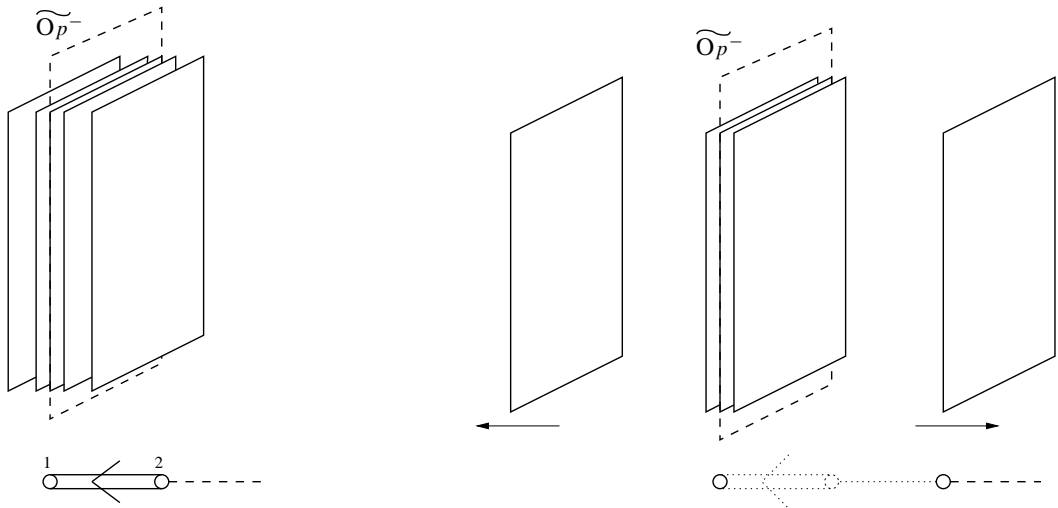}
\caption{$SO(2N+1)$: the short root should be identified with the
orientifold plane.}
\end{center}
\end{figure}

We can now explain the absence of a Higgs branch for a \D{0}-brane
near an \Omt{4} plus $1$ pair of D4-branes. Suppose we consider
\Omt{4} with 2 coincident pairs of D4-branes; there is an $SO(5)$
spacetime gauge symmetry. Approach the configuration with a
\D{0}-brane and let it fatten into an instanton. If we break the
group to $SO(3)$ by moving away a pair of D4-branes, what happens
to the instanton? It cannot remain embedded in the $SO(3)$
subgroup since that requires at least charge $2$. Therefore, it
must become a small instanton, which is a D0-brane again.

We can extend these remarks for bound states of $k$ D0-branes
probing \Omt{4} with a pair of D4-branes. The ADHM setup
of~\cite{Christ:jy} again involves $4k-3$ parameters, instead of
the $8k-3$ one would expect if $k$ is to be identified with the
instanton charge. While it might appear this construction is
flawed, we know that it has a physical realization in terms of
branes, and we now understand the resolution of this paradox.

If $k$ is even, the \D{0}-branes can dissolve in the combined D4,
\Omt{4} system. The actual instanton charge is $k' =k/2$, and the
moduli space is described by the expected $8k'-3$ parameters. If
$k$ is odd, then an even number of D0-branes can dissolve in the
\D{4}, \Omt{4} system. The maximal instanton charge is $k' =
(k-1)/2$, and the moduli space is described by $8k'-3 = 4k-7$
parameters. The remaining 4 parameters have a natural
interpretation as the coordinates of the remaining \D{0}-brane.
It can be viewed as an instanton, but one with necessarily zero
size.

\section{Twisting the $E_8$ Tensor Theory}
\subsection{Three-dimensional superconformal theories}
\label{scft} Our starting point is the $E_8\times E_8$ heterotic
string compactified on $T^3$. As explained
in~\cite{deBoer:2001px}, there are $6$ distinct components
preserving all $16$ supersymmetries in the moduli space of the
compactified string. Each component is distinguished by the
choice of $E_8\times E_8$ gauge bundle. The gauge bundle in each
component is flat, but characterized by a fractional Chern-Simons
(CS) invariant~\cite{Borel:1999bx}\ defined by,
 \be CS(A) = {1\over 16 \pi^2 h}
\int \tr(AdA + {2\over 3} A^3), \ee with $h$ the dual Coxeter
number.
 These distinct string
compactifications exist because $E_8$ gauge theory has non-trivial
`triples' of commuting connections. We can label the triples by
an abelian group, $\Z_m$, where $m=1,\ldots 6$. For a $\Z_m$
triple, the CS invariant can take values ${p\over m} \,{\rm
mod}\,\Z$ with $p$ and $m$ relatively prime $$ (p,m)=1.$$ There
are, therefore, $12$ components in the gauge theory moduli space.
To construct a string compactification, we take $\Z_m$ triples for
both $E_8$ factors. In one $E_8$ factor, we embed CS invariant
${1\over m} \, {\rm mod}\, \Z$ while in the other factor, we
embed CS invariant $-{1\over m} \, {\rm mod}\, \Z$. Anomaly
cancellation requires that the total CS invariant vanish.

What is important for us is that novel non-simply-laced gauge
groups arise in these new sectors, generalizing the structure
that first appeared in the CHL string~\cite{Chaudhuri:1995fk}. We
naturally ask: what is the small size limit of instantons in such
groups? Via standard arguments, we expect the spacetime gauge
symmetry to become a global symmetry of the resulting
$3$-dimensional theory. In table~\ref{table:symmetry}, we list
the maximal enhanced global symmetries that can arise in the
theories describing the zero size limit of these instantons.
Included in the table is the number of light hypermultiplets on
the Higgs branch, including the free hypermultiplet parametrizing
the center of mass of the instanton. The counting of
hypermultiplets follows from the index theorem on $\R^4$ which
gives the dimension of the moduli space $\M_N (G)$ of $N$
instantons in a group $G$, \be\label{indexformula} {\rm dim} \,
\M_N(G) = 4 h(G) N, \ee where $h(G)$ is the dual Coxeter number.
This formula is only valid for a sufficiently large instanton
number, which is $2$ for $G_2, F_4$, $3$ for $E_8$ and $4$ for
$Sp(4)$~\cite{Bernard:nr}. However, we can always start with a
gauge bundle with sufficiently large instanton number, and shrink
one instanton. This requires tuning $4h(G)$ parameters, giving us
the count of hypermultiplets. The Higgs branch for $N$ small
instantons (when there is a non-trivial branch) then describes
the moduli space of $N$ instantons embedded in the maximal global
symmetry group listed in table~\ref{table:symmetry}.

\begin{table}
\begin{center}
\renewcommand{\arraystretch}{1.5}
\begin{tabular}{|c|c|c|c|} \hline
Triple  & Maximal global symmetries  & Hypermultiplets  \\  \hline
$\Z_1$ & $E_8$ & $30 N$   \\ \hline $\Z_2$ & $F_4, C_4$ & $9N$, $5N$  \\
\hline $\Z_3$ & $G_2$ & $4N$  \\ \hline $\Z_4$ & $A_1$ & $2N$ \\
\hline $\Z_5$ & $\{ e\} $ & 1 \\ \hline $\Z_6$ & $\{ e\} $ & 1 \\
\hline
\end{tabular}
\renewcommand{\arraystretch}{1.0}
\caption{Some properties of the small instanton theories for the $\Z_m$ triples.
 } \label{table:symmetry}
\end{center}
\end{table}

Included in the superconformal field theories that we find are
some with exotic $G_2$ and $F_4$ global symmetries. These differ
from the probe theories found in~\cite{Aharony:1996bi}\ which are
$4$-dimensional theories with $4$ rather than $8$ supercharges.
Indeed, in our construction, it appears that these exotic global
symmetries do not appear above three dimensions. These theories
also seem to differ from compactifications of the $6$-dimensional
interacting theories found in~\cite{Intriligator:1997dh,
Blum:1997mm}\ which can have exotic local gauge symmetries.

\subsection{A tensor theory interpretation}

How are we to interpret these new small instantons? To answer this question, we return to a
more familiar case of type I compactified on $T^2$. The non-perturbative gauge group for the
type I string is  $Spin(32)/\Z_2$. We need to choose a flat  $Spin(32)/\Z_2$ bundle on $T^2$
to specify our compactification. Such a bundle is described by two holonomies
$\Omega_1$, and $\Omega_2$.  Flatness of
the bundle implies that the two holonomies commute in
$Spin(32)/\Z_2$. Lifting the holonomies $\Omega_{1,2}$ to elements
$\widetilde{\Omega}_{1,2}$ in $Spin(32)$ results in the following
lift of the commutation relation,
\be
\widetilde{\Omega}_1 \widetilde{\Omega}_2 \widetilde{\Omega}_1^{-1}
\widetilde{\Omega}_2^{-1} = z.
\ee
Here $z$ is either the identity, or the generator of the $\Z_2$ defining
$Spin(32)/\Z_2$. The element $z$ can be non-trivial because the
representations that are sensitive to this $\Z_2$ are absent in
the theory. The smallest of these representations is the vector representation.
Bundles where $z$ is trivial have vector structure (VS), while bundles where $z$ is
non-trivial are without vector structure (NVS).

Let us consider a stack of N D5-branes wrapping the torus. The
type I D5-branes support an $Sp(N)$ gauge
group~\cite{Witten:1995gx}. The combined D9-D5 system has a gauge
group whose simply connected cover is $Spin(32) \times Sp(N)$.
The 5-9 and 9-5 strings stretching between the two sets of branes
give states transforming in the $(\mathbf{32}, \mathbf{2N})$. In
a NVS compactification, this is inconsistent because it involves
the $\mathbf{32}$ vector representation.

The resolution of this paradox, proposed in~\cite{Witten:1997bs},
is to note that the vector representation is accompanied by the
$\mathbf{2N}$ of $Sp(N)$. Since the 5-5 strings only give
representations of $Sp(N)/\Z_2$, we see that it is possible to
choose a bundle for on the D5-branes on $T^2$ with commuting
holonomies in $Sp(N)/\Z_2$ that cannot be lifted to commuting
holonomies  in $Sp(N)$. In fact, choosing this twisted $Sp(N)$
bundle appears to be the only way to resolve the problem with the
5-9 and 9-5 strings. By choosing such a bundle, the $\mathbf{32}$
at one string end-point picks up a non-trivial phase when
transported around certain closed curves, but this is always
cancelled by an identical phase picked up by the $\mathbf{2N}$ at
the other end of the string.\footnote{Another easy way to see that
this must happen is to apply one T-duality to one of the circles
of the $T^2$. A discrete $B$-field present on the $T^2$ (and
correlated with NVS~\cite{Bianchi:1991eu, Sen:1997pm}) will
result in a dual theory on a M\"obius
strip~\cite{Keurentjes:2000bs} The twisted bundle is the lowest
energy solution to the requirement that the D8 and D4-branes
close on this space.} Note that the actual D9-D5 gauge group is
therefore $(Spin(32) \times Sp(N))/\Z_2$. Here, $\Z_2$ is the
diagonal subgroup formed from the product of the $\Z_2$ of
$Spin(32)/\Z_2$ and the $\Z_2$ center of $Sp(N)$.

In the NVS sector, $SU(2)$ gauge groups at level 1 and level 2.
Both levels are possible because the NVS compactification
involves a projection on $Spin(32)/\Z_2$. Surviving $SU(2)$
subgroups are either invariant under this projection (resulting
in level 1 subgroups), or are the diagonal combination of two
$SU(2)$ subgroups which are exchanged by the projection. This
latter case is a level 2 subgroup. Instantons can be embedded in
either kind of group. However, the smallest charge for a `level
1' instanton is one-half the minimal charge of a `level 2'
instanton.

In the small size limit, these instantons become BPS $5$-branes
with tensions proportional to their charges. However, a level 1
5-brane cannot fatten into an instanton of a level 2 group
because its charge is too small. The only possible Higgs branch
for such a $5$-brane corresponds to fattening into a level 1
instanton. However, level 1 subgroups are only possible in a
limited locus of the moduli space. By tuning the Wilson lines, we
can break all level 1 subgroups to abelian subgroups. In this
case, there is no Higgs branch at all.

The lesson we take from this example is that NVS, which is a
property of the string compactification, correlates with `no
symplectic structure,' which is a property of the gauge theory on
the D5-brane. Indeed, it is precisely the one non-trivial choice
of 't Hooft twist. From the results of~\cite{deBoer:2001px}, we
know that for a $T^3$ compactification, the $\Z_2$ triple and NVS
are in the same component of the moduli space. A T-duality relates
both descriptions. This same T-duality maps the D5-brane to an
$E_8$ 5-brane. It is therefore natural for us to conjecture that
the $E_8$ tensor theory has a 3-cycle analogue of an 't Hooft
twist.

We need to determine how many distinct twists are possible on
$T^3$. The structure of the instanton depends only on the gauge
bundle for one of the $E_8$ factors. Global considerations, like
anomaly cancellation between $E_8$ factors, are unimportant. To
each triple of $E_8$, we should then expect a distinct $p\over m$
twist in the $E_8$ tensor theory, where $p\over m$ specifies the
CS invariant of the ambient $E_8$ bundle with $\Z_m$ triple. Of
these $12$ twisted sectors, those associated to the $\Z_2, \Z_3$
and $\Z_4$ triples have Higgs branches because there is
non-abelian spacetime gauge symmetry. For the $\Z_5, \Z_6$ cases,
there is no enhanced gauge symmetry\footnote{Excluding possible
enhanced gauge symmetries from Kaluza-Klein gauge bosons.} so the
5-brane cannot fatten.

\subsection{Coulomb branches from duality}

So far, we have described the Higgs branch for each twist of the
$E_8$ theory. Now we turn to the Coulomb branch. Duality will aid
us in understanding the structure of the Coulomb branch. Let us
start with $N$ D5-branes on $T^2$ with NVS. The structure of
$Sp(N)/\Z_2$ bundles can be studied for each choice of
$N$~\cite{Schweigert:1996tg} (see
also~\cite{Witten:1997bs,Keurentjes:2000bs}). Take $N=1$ to
start. The moduli space of twisted $Sp(1)\sim SU(2)$ bundles on
$T^2$ has one component. The moduli space consists of a single
point. We can be quite explicit here, and take for the holonomies
$\Omega_1$ and $\Omega_2$ on $T^2$ \be \Omega_1=i\sigma^2, \qquad
\Omega_2=i\sigma^3 \ee where $\sigma^i$ are the usual Pauli
matrices. These two holonomies clearly anti-commute. The gauge
group left unbroken by this compactification is the $\Z_2$
generated by the center of $SU(2)$.

 This moduli space is nicely interpreted in the language of
orientifolds~\cite{Keurentjes:2000bs}. T-dualizing along both
cycles of $T^2$ leads to the following configuration of
$O7$-planes $(+,-,-,-)$. We use the conventions
of~\cite{deBoer:2001px}\ where $+$ refers to an \Op{7}-plane, $-$
refers to \Om{7}, while $-'$ refers to an \Omt{7} if there were
such a plane. The D5-brane turns into a D3-brane at a point on
$T^2$. What corresponds to twisting the $Sp(1)$ bundle is
sticking the D3-brane at the position of the $O7^+$ orientifold.

There is a final dual description of this moduli space that will
be useful for us. This description is given by F theory
compactified on a $K3$ surface with a frozen $D_8$
singularity~\cite{Witten:1997bs}. In this description, the
wrapped D5-brane becomes a D3-brane trapped at the frozen $D_8$
singularity. If we increase $N$, the structure becomes more
interesting. For $N$ even, there are $N$ moduli for a twisted
$Sp(N)/\Z_2$ bundle. However, for $N$ odd, there are only $N-1$
moduli. Adding an additional D5-brane to an even number of branes
has no effect on the number of moduli. In the T-dual language,
the means that an even number of D3-branes can wander over the
torus, but an additional D3-brane is always bound to the
$O7^+$-plane. This gives us the structure of the Coulomb branch
of $N$ D5-branes on $T^2$.

Replacing branes on $T^2$ by branes wrapping $T^3$ requires
specifying the holonomy for the twisted bundle around the extra
circle. Let us return to $N=1$ where the holonomy, $\Omega_3$,
must be $\pm \bf{1}$. There are therefore $2$ components in the
moduli space. The resulting $2+1$-dimensional gauge group is
still $\Z_2$ so there is no further modulus from dualizing the
photon. In a conventional untwisted compactification, the photon
would otherwise give rise to an additional circle modulus.  The
two components, each consisting of a point, have CS invariants,
$0$ and $1/2$ mod $\Z$.

The T-dual orientifold configuration is $(+^2, -^6)$, and the two
components correspond to sticking the D2-brane that results from
T-dualizing the D5-brane at either $O6^+$-plane. Note that there
is a well-defined moduli space for a D2-brane so there is no
subtlety in discussing its position. The last dual description
involves M theory on a $K3$ with $2$ frozen $D_4$ singularities.
We see that a frozen $D_4$ singularity traps a single M2-brane.
For higher $N$, the physics is different. For $N$ even, there are
two distinct components in the moduli space. Either all pairs of
M2-branes wander freely over the $K3$ surface, or there can be a
single M2-brane bound to each $D_4$ singularity.

\begin{figure}[h]
\begin{center}
\includegraphics[width=15cm]{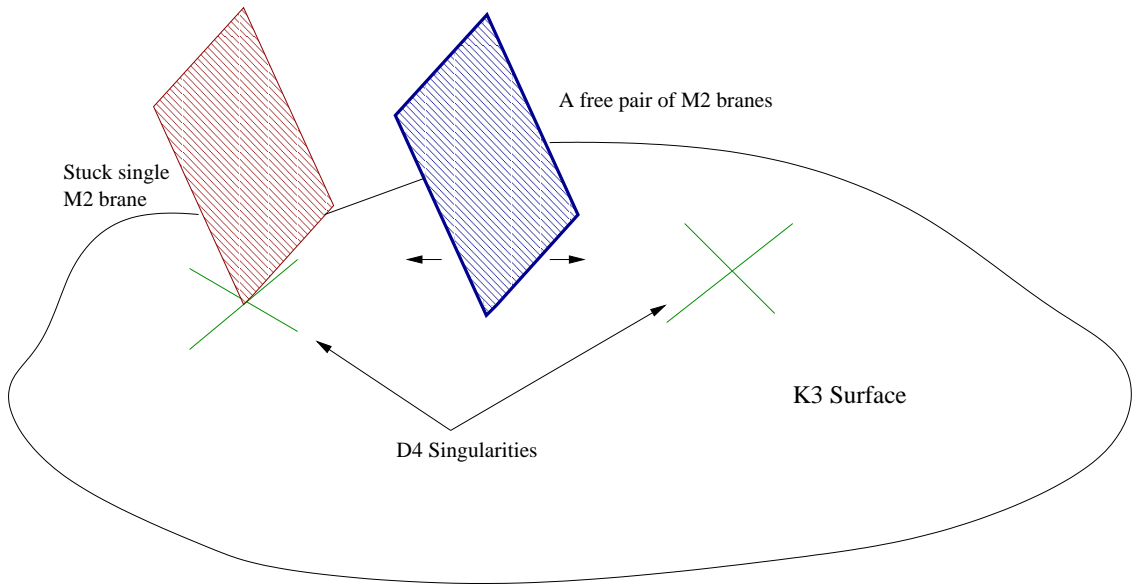}
\caption{A depiction of the Coulomb branch for D5-branes wrapping
$T^3$.}
\end{center}
\end{figure}

For $N$ odd, there are two isomorphic components in the moduli
space. One M2-brane must bind two either of the $D_4$
singularities. The remaining branes wander in pairs over the $K3$
surface. This structure, arrived at by considering twisted
$Sp(N)$ bundles on $T^3$, confirms a result found by a quite
different argument in~\cite{Atiyah:2001qf}: namely, that a
membrane near a frozen $D_{4}$ singularity corresponds to
instanton charge $2$ ($N=2$). We see quite explicitly that a
charge $1$ instanton corresponds to a membrane trapped at the
singularity.

This dictionary between M2-branes at frozen singularities, and
the Coulomb branch of twisted wrapped 5-branes is quite critical
for us. We shall use consistency with the results
of~\cite{deBoer:2001px, Atiyah:2001qf}\ to predict the Coulomb
branch of the twisted $E_8$ tensor theory. The structure of the
Coulomb branch can be described as follows: let us take $N$ $E_8$
5-branes wrapping $T^3$ with a $p\over m$ twist. Let us further
assume that the spacetime gauge symmetry is abelian. The dual
description is M theory on a $K3$ with a pair of frozen $D_4$
singularities for $m=2$, $E_6$ singularities for $m=3$, $E_7$
singularities for $m=4$, and $E_8$ singularities for $m=5,6$. The
value of $p$ determines the flux of the M theory $3$-form through
the singularity. Consistency with~\cite{Atiyah:2001qf}\ requires
that a wandering membrane correspond to instanton charge $m$. We
therefore split $N=p+p'$ where $p$ M2-branes are localized at one
singularity, while $p'$ M2-branes live at the other singularity.
If either $p\geq m$ or $p'\geq m$ then $m$ branes can leave the
singularity and wander into the bulk.

There is one final point that merits comment. For untwisted $T^3$
compactifications of $E_8$ 5-branes, which give theories with
global $E_6,E_7$ and $E_8$ symmetries, there exist mirror
realizations. In the mirrors, the moduli space of instantons is
realized on the Coulomb rather than the Higgs branch. These mirror
descriptions can be constructed as the IR fixed points of the
conventional $E_6,E_7$ and $E_8$ quiver gauge
theories~\cite{Intriligator:1996ex}. For our twisted
compactifications, the mirrors are currently unknown. It is not
hard to see that in the IR, they correspond to the theories on
coincident M2-branes localized at (partially) frozen $D_{4+n},
E_6, E_7$, and $E_8$ singularities. Depending on the choice of
flux through the singularity, the global symmetry will correspond
to the symmetry of table~\ref{table:symmetry}. Further the
Coulomb branch will correspond to the appropriate moduli space of
instantons, while the Higgs branch will correspond to the Coulomb
branch described just above. Whether these theories can be given
Lagrangian descriptions is an outstanding question.

\subsection{Testing string/string duality}

At this stage, we cannot resist applying our earlier results to
string/string duality. Let us turn to the case of 5-branes
wrapping $T^4$. Five-branes give rise to strings which we can
interpret as fundamental strings of a dual description. The
standard duality is between heterotic/type I on $T^4$ and type
IIA on $K3$. Consider a IIA string wrapping the circle of
$K3\times S^1$. The massless modes of the string correspond to
states with $L_0= \bar{L}_0=0$, and there are $24$ such states
corresponding to the cohomology of $K3$.

Let us recall how these light modes arise from a type I dual
description. These states then correspond to light modes of a
D5-brane wrapping $T^4\times S^1$. After T-dualizing on all
circles, this system becomes $1$ D0-brane together with a
collection of $32$ \Om{4}-planes and $16$ pairs of D4-branes. The
D0-brane sees $T^5/\Z_2$ as its moduli space. By correctly
counting gauge invariant ground states, we see that there are $8$
light modes from motion on the moduli space~\cite{Witten:1995gx}.
There are no additional ground states from the D0-brane near an
\Om{4}-plane. These correspond to ground states in a pure $Sp(1)$
gauge theory with $8$ supercharges, but there are no such
states~\cite{Sethi:1997pa, Yi:1997eg}. However, there is
precisely one ground state from the D0-brane near each
D4-brane~\cite{Sethi:1996kj}. This gives the required $16$
additional light modes.

The prior discussion assumes that the type I gauge bundle on $T^4$
is in the component of the moduli space containing the trivial
Wilson line. It is natural to extend the analysis to the case of
a NVS compactification. The dual description is IIA on a $K3$
surface with $8$ frozen $A_1$
singularities~\cite{Schwarz:1995bj}. This is not really a
perturbative string compactification because of the presence of
torsion RR 1-form fluxes. However, it is essentially clear that a
string wrapping an extra circle has $16$ rather than $24$ $L_0=
\bar{L}_0=0$ light modes. The dual orientifold description now
contains $24$ \Om{4}-planes, $8$ \Op{4}-planes, and $8$ D4-branes
(see~\cite{deBoer:2001px}\ for a discussion). From this
description, we can determine the light degrees of freedom:
again, there are $8$ modes from motion of the D0-brane on
$T^5/\Z_2$, no modes from the \Om{4}-planes and $8$ modes from
the D4-branes. Using our result from
section~\ref{Boundstatecomp}, we see that we also obtain no modes
from the \Op{4}-planes. In total, we obtain precisely $16$ modes,
which is in accord with generalized string/string duality.

There is one more point worth stressing about IIA on a $K3$ with
$8$ frozen $A_1$ singularities. As in all the cases we have
considered, the non-simply-laced groups that appear in the
low-energy theory imply the existence of fractional instantons.
What is the stringy description for the small size limit of these
instantons? As we have argued, a small (fractional) instanton in
the dual heterotic theory on $T^4$ is a $5$-brane wrapping $T^4$.
In the dual picture, these solitonic strings become fundamental
strings of IIA. The duality dictionary therefore implies
fractional strings localized at singularities. In this way, we
rediscover the fractional strings of~\cite{Schwarz:1995bj}. This
argument, which only uses low-energy physics, complements the
reasoning of~\cite{Schwarz:1995bj} which appeals to M theory in
order to predict fractional strings.

\subsection{Quadruple and NVS compactifications}

As a final topic in string/string duality, we will now briefly
visit the case of $Spin(32)/\Z_2$ string theory on $T^4$. We will give
a new argument for the equivalence
of an NVS compactification with a quadruple compactification. A
perturbative argument based on T-duality appears
in~\cite{deBoer:2001px}. Let us recall that a quadruple
compactification corresponds to a gauge bundle in a disconnected
component in the moduli space of flat connections on $T^4$.
However, all possible CS invariants vanish. In particular,
$Spin(32)/\Z_2$ admits a quadruple compactification.

To construct a quadruple, we first turn on Wilson lines on 2 of
the 4 cycles available on $T^4$, such that a non-simply-connected
group remains unbroken. According to the analysis by
Schweigert~\cite{Schweigert:1996tg}, the twisted bundles with
non-simply-connected structure groups on $T^2$ are related to
diagram automorphisms of the \emph{extended} Dynkin diagram. The
groups we will encounter are of the $SO(4N)$ type; we therefore
study the diagram automorphism of the extended diagram
corresponding to $D_{2n}$.

\begin{figure}[h]
\begin{center}
\includegraphics[width=10cm]{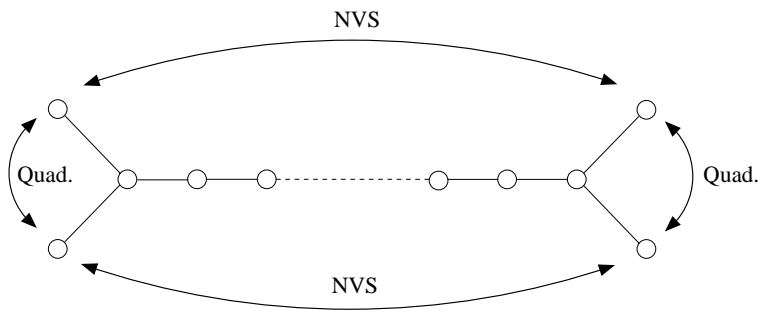}
\caption{The automorphisms which act on the extended Dynkin
diagram of $D_{2n}$, and their correspondence with the quadruple
and NVS compactifications.} \label{Dnext}
\end{center}
\end{figure}

There are two automorphisms that are relevant to us; see
figure~\ref{Dnext}). We can flip the extended diagram from left to
right, or up/down. The first corresponds to the no vector
structure compactification; the second one gives the quadruple
compactification.\footnote{The twist of one of the two ``forks''
of the extended Dynkin diagram corresponds to the outer
automorphism of $D_{2n}$.}

Start with type I on $T^4$. Turn on 2 Wilson lines on 2 cycles
such that the gauge group is broken to a manifest $O(8)^4$. From
duality with the heterotic $Spin(32)/\Z_2$ string, it is not hard
to deduce that the actual group is $Spin(8)^4/(\Z_2 \times \Z_2)$.
The $\Z_2 \times \Z_2$ factor appearing in the quotient is the
diagonal product of the centers of the 4 $Spin(8)$ factors. The
two $\Z_2$ factors are related to the two diagram automorphisms we
mentioned earlier.

Let us turn on a third Wilson line that breaks the (manifest)
gauge group to $O(4)^8$. T-dualizing each direction of $T^3$ that
carries a non-trivial Wilson line leads us to type IIA on
$T^3/\Z_2 \times S^1$. There are $8$ \Om{6}-planes each with $2$
pairs of coincident D6-branes. This lifts to M theory on
$T^4/\Z_2 \times S^1$. We now have two interesting possibilities
for the Wilson line around the extra circle: the first is to
impose a holonomy on $S^1$ that exchanges the $O(4)$ groups in
pairs. This reduces the rank by $8$, and results in an NVS
compactification. The second possibility is to impose a holonomy
that breaks each $O(4) \rightarrow O(3)\times O(1)$. This gives a
quadruple compactification, and also reduces the rank by $8$.

\begin{figure}[h]
\begin{center}
\includegraphics[width=10cm]{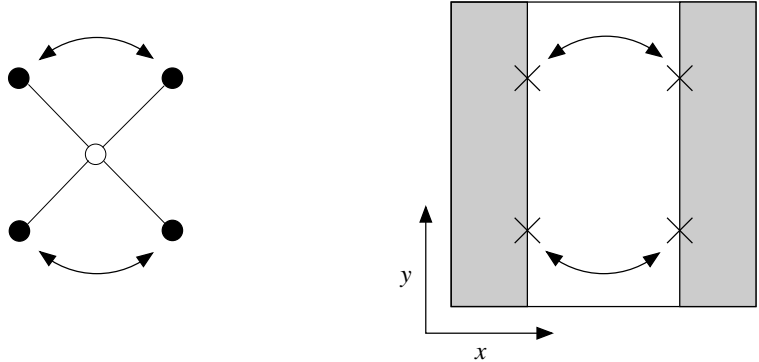}
\caption{Geometry of the duality between NVS and the quadruple
compactification. Left: the extended Dynkin diagram of $D_4$.
Right: a fundamental domain for $(\R^2 \times T^2)/\Z_2$.}
\label{dualnvsquad}
\end{center}
\end{figure}

Let us consider these theories after T-dualizing each cycle of
$T^3$. It is sufficient for us to focus on $2$ \Om{6}-planes and
their associated $4$ pairs of D6-branes. In general \Om{6} with
D6-branes lifts to a $D$-type singularity in M theory. In this
case where there are only $2$ pairs of D6-branes, we obtain
locally $4$ $A_1$ singularities, or $2$ ``$D_2$'' singularities.
The explicit geometry in this case is $(R^2 \times T^2)/\Z_2$.
Imposing the first flavor of Wilson line corresponds to
exchanging pairs of $A_1$ singularities on traveling around the
$S^1$. How about the second case? Reducing $O(4)\rightarrow
O(3)\times O(1)$ requires a holonomy on $S^1$ with determinant
$-1$. Conjugating by this holonomy is equivalent to acting by an
outer automorphism, and so also corresponds to exchanging $2$
$A_1$ singularities in the M theory lift. However, the exchanged
$A_1$ singularities now sit in the same fiber!

The M theory geometry is both simple and pretty, and is depicted
in figure~\ref{dualnvsquad}. On the left, we have the diagram of
$D_4$, corresponding to $Spin(8)$. Breaking this to $O(4)^2$
corresponds to erasing the middle node. This should be compared
with the right hand side, where we have depicted the $T^2/\Z_2$
which appears in $(\R^2 \times T^2)/\Z_2$. The singularities,
marked by crosses, just correspond to the black nodes of the
Dynkin diagram on the left. The white node is the $T^2/\Z_2$
(which is topologically a 2-sphere) intersecting these
singularities. The exchange of singularities depicted produces
either an NVS or a quadruple compactification; it just depends
whether on the right hand side we interpret the $x$ or the $y$
direction as the ``eleventh'' M theory direction. In other words,
these two cases are related by a suitable $9-11$ flip. This
provides a strong coupling version of the argument given
in~\cite{deBoer:2001px}.

\section{Domain Walls in Heterotic/Type I String Theory}

The final topic that we shall discuss are domain walls in
heterotic and type I string theory. In D=8 dimensions, there are
two distinct $T^2$ type I/heterotic compactifications: the NVS
compactification and the standard compactification. The gauge
bundles are topologically distinct. Therefore there is no field
theoretic domain wall that can smoothly interpolate between these
two distinct vacua. However, as shall describe, there is a
stringy domain wall that interpolates between these two vacua.

In $D=7$, there are many components in the heterotic string
moduli space on $T^3$. These components are distinguished by the
CS invariants of their $E_8\times E_8$ gauge bundles. The
situation is nicer than in $D=8$ because the gauge bundles are all
topologically trivial. As a starting point, we can then consider
pure $E_8$ gauge theory on $T^3\times \R$. Let us parametrize
$\R$ by a coordinate $x\in (-\infty, \infty).$ Consider an
instanton configuration that interpolates between an $E_8$ gauge
bundle with CS invariant,\footnote{The possible CS invariants for
$E_8$ are given in section~\ref{scft}.} $CS(-\infty)$ at
$x=-\infty$ to one with CS invariant, $CS(+\infty)$ at
$x=\infty$. Such a BPS instanton is (generically) fractionally
charged, satisfying: \be \frac{1}{8 \pi ^2 N_{R}}
\int~\textrm{Tr}\,(F \wedge F) = CS(\infty) -
CS(-\infty).\label{instbound}\ee It is clear that there are
smooth gauge-field configurations in a sector with fixed
instanton charge. For example, a smooth interpolation between the
flat gauge-field configurations at $x=\pm \infty$ works. What has
yet to be shown is that {\it smooth} solutions exist which
saturate the BPS bound~\C{instbound}. The existence of such
solutions might be demonstrated by extending the methods
of~\cite{Jardim:2001cw, Ford:2000zt}. It might even be possible
to explicitly construct such a solution using the connections
worked out in~\cite{Selivanov:2000kg}. This is an important issue
in field theory, but less critical in string theory. A non-BPS
configuration will decay down to something reasonable in string
theory, and cannot decay away entirely because of its charge.

\begin{figure}[h]
\begin{center}
\includegraphics[width=15cm]{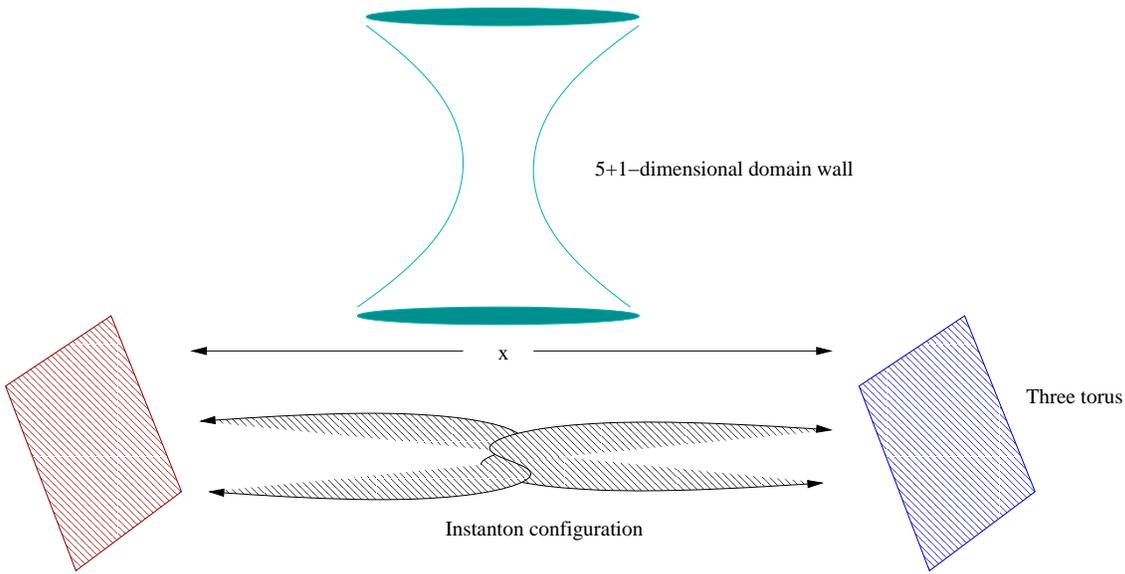}
\caption{The embedded instanton/anti-instanton pair interpolate
between fixed asymptotic CS invariants, giving a space-time
domain wall.} \label{domain}
\end{center}
\end{figure}

When embedded in the heterotic string, we want to satisfy anomaly
cancellation as $|x|\rightarrow \infty$ with $H_3=0$. From the
usual relation, \be H_3 = dB_2 + CS(\omega) - CS(A), \ee where
$\omega$ is the spin connection, and $A$ the gauge-field, we see
that we must embed a fractional instanton in one $E_8$ factor,
and a fractional instanton of opposite charge in the second
factor. The total configuration is then not BPS, although the
instanton in each factor can be BPS. Since both vacua at
$|x|=\infty$ are supersymmetric with zero cosmological constant,
we do not expect static domain wall solutions when we couple
gravity to our $E_8\times E_8$ Yang-Mills theory. Rather, the
domain wall inflates. For a review of domain walls in
supergravity, see~\cite{Cvetic:1996vr}.

\subsection{Domain walls from orientifold planes}

We start by constructing a domain wall for the $D=8$ case.
Consider a stack of NS5-branes at a point on a four manifold,
$M_4$. We can measure the 5-brane charge $k$ by computing \be
\int_{\partial M_4} H_3 = k. \ee We want to introduce an
orientifold plane in such a way that the NS5-branes are inside
the world-volume of the plane. The orientifold projection inverts
$B_2$, and therefore $H_3$. If we want to allow NS5-branes, we
need to accompany the orientifold action with a geometrical
action on the transverse space.\footnote{A $(-1)^{F_L}$ which is
also needed in some dimensions does not change this argument
since it has no effect on $B_2$.} This means that the orientifold
plane has to be an O6 or an O8-plane. The case of an O6-plane is
well-known, and has been studied in many
configurations~\cite{Evans:1997hk, Hanany:2000fq,
Bertoldi:2002nn}. The case of interest for us is the \Om{8}-plane.
Although this plane does not make sense in isolation i.e. without
additional D8-branes, we will soon make our local construction
globally sensible.

Let us take our space $M_4$ to be $\R^4$. The \Om{8}-plane
intersects the $S^3=\partial \R^4$ on an $S^2$. We can now turn
the $H_3$ integral into a $B_2$ integral over the boundary of the
orientifolded $3$-sphere. \be \int_{\partial (\R^3\times
\R/\Z_2)} H_3 = \int_{\partial (\R^3\times \R/\Z_2)} \dif B_2 =
\int_{\Om{8} \bigcap S^3} B_2 = k/2. \ee In particular, when $k$
is odd, the \Om{8}-plane supports a half-integer $B_2$-flux on its
world-volume (the same conclusion was reached in
\cite{Bachas:2001id}).

Things are more interesting when the space transverse to the
5-branes is partially compactified. In particular, let us take $k$
NS5-branes at a point on $T^2 \times \R^2$. We introduce an
\Om{8}-plane which wraps the $T^2$.  The orientifolded space is
now $(T^2 \times \R) \times \R/\Z_2$. Any transverse component of
the $B_2$-field can be gauged away (provided the background is
flat). We now repeat the previous argument, and compute \be
\int_{\partial (T^2 \times \R\times \R/\Z_2)} H_3 =
\int_{\partial (T^2 \times \R\times \R/\Z_2)} \dif B_2 =
\left(\int_{T^2} B_2 \right)( \infty ) - \left(\int_{T^2} B_2
\right)( -\infty) = k/2, \ee where $\pm \infty$ refers to the
boundary of $\R$. We see then that the stack of NS5-branes forms a
domain wall. If we traverse the wall, the $B_2$-flux through the
$T^2$ jumps by $k/2$ units. The most interesting case is when $k$
is odd. Necessarily, on one side of the domain wall we have an odd
$B_2$-flux, while on the other side, we have an even $B_2$-flux.
If there are $D8$-branes coincident with the \Om{8}-plane, we can
now invoke the arguments of \cite{Bianchi:1991eu, Sen:1997pm}. On
one side of the domain wall,  we have vector structure on the
two-torus, while on the other side we do not. Note that it is
absolutely crucial for this argument that the NS5-branes are
effectively codimension 1 with respect to the \Om{8}-plane.
Otherwise, we do not have a domain wall. Studying the effect of
T-duality on this configuration should be interesting.

We will take a different tact, and replace $\R/\Z_2$ by
$S^1/\Z_2$. Forgetting the NS5-branes for the moment, the
resulting configuration is type I' on $S^1/\Z_2 \times T^2$. With
a single T-duality, we could arrive at type I on $T^3$. We,
however, are interested in the M theory description of this setup
so we perform a 9-11 flip, compactifying from M theory to IIA on
$S^1/\Z_2$. This gives the Horava-Witten theory on a
three-torus~\cite{Horava:1995qa}. The NS5-branes become heterotic
M theory 5-branes, while the $B_2$ fluxes become $C_3$ fluxes.

Combined with the arguments given in~\cite{deBoer:2001px}, and the
connection between 5-branes and small instantons in heterotic
string theory, we are inevitably lead to the conclusion that we
have found a domain wall. It interpolates between two $E_8$
vacua: one is the standard $T^3$ compactification, while the
other is a $\Z_2$ triple. When crossing the interval between the
two HW walls, the $C_3$-field cannot jump. This is achieved by
embedding a second domain wall in the other HW wall. This
configuration corresponds precisely to the fractional instanton
solution that we expected to find.

 In the strong coupling limit,
the HW walls are far apart, and the $E_8$ groups are localized at
each wall. In the weak coupling limit, when the HW walls come
close together, the picture of localized gauge groups is no longer
accurate, but happily, the picture of localized 5-branes cannot
be trusted either. Instead it seems likely that the two
oppositely charged 5-branes form a non-BPS configuration that is
smeared along the interval between the 5-branes. Although there
seems to be no direct link, a parallel with the behaviour of the
non-BPS  string of type I' as a function of the distance between
the $\Om{8}$-planes~\cite{Dasgupta:1999hw}\ is very suggestive.

\subsection{Beyond $T^3$}

Much of our discussion should extend to lower-dimensional
compactifications. Type I on $T^4$ has a number of distinct vacua
again distinguished by the choice of gauge bundle. Of particular
interest is the quadruple compactification. This bundle cannot be
smoothly deformed away, yet there is no topological class that
characterizes the bundle. It is natural for us to conjecture that
Yang-Mills theory on $T^4\times \R$ has finite action field
configurations that interpolate between the quadruple and the
trivial vacuum. There is no natural candidate for a quantum
number that can characterize this configuration so we expect it
to be non-BPS. When embedded in string theory, the domain wall
corresponds to a non-BPS $4$-brane. Interpreting the domain wall
from the perspective of the string worldsheet is quite intriguing
because it involves unwinding a discrete RR 4-form
flux~\cite{Keurentjes:2001cp}. There will also be cases where NVS
compactifications are combined with quadruples to give new kinds
of domain walls.

The situation is similar for $T^5$ where there exist quintuple
compactifications of the $E_8\times E_8$ string. Again there
should be concomitant domain wall solutions, which now correspond
to non-BPS $3$-branes. In this case, we can embed a quintuple in
either $E_8$, or in both factors, so there are at least two
distinct domain walls.

{}For type I on $T^6$ (or any $6$-dimensional Calabi-Yau
compactification), something new should happen. There are $2$
classes of compactification which are distinguished by a discrete
NS-NS $B_6$ flux~\cite{Morrison:2001ct}. As we approach a domain
wall, which will be a membrane in the $4$ space-time dimensions,
the field strength $H_7=dB_6$ will vary. Note that $H_7$ is
supported in the $6$ internal space coordinates, and along the
spatial direction transverse to the domain wall. However, $H_3 =
\ast H_7$, so from the perspective of a fundamental string, there
is a varying electric flux. This suggests the intriguing
possibility that space-time non-commutativity might be involved.

\section{Acknowledgements}
It is our pleasure to thank P. van Baal, N. Dorey, A. Hanany, A.
Hashimoto, K. Hori, S. Ross, E. Witten, and P. Yi for helpful
discussions. We would also like to thank the organizers of the
Amsterdam 2001 Summer Workshop where we initiated this project.
The work of A.~K. is supported in part by EU contract
HPRN-CT-2000-00122, while the work of S.~S. is supported in part
by NSF CAREER Grant No. PHY-0094328, and by the Alfred P. Sloan
Foundation.

\appendix
\section{Symplectic Group Actions}

We will summarize some useful relations between quaternions and
symplectic groups. Let us label a basis for our quaternions by
$\{ {\bf 1} , I, J, K\}$ where, $$I^2=J^2=K^2=-{\bf 1}, \qquad IJK =
- {\bf 1}. $$ A quaternion $q$ can then be expanded in components
$$ q = q^1 + I q^2 + J q^3 + K q^4. $$ The conjugate quaternion
$\bar{q}$ has an expansion $$ \bar{q} = q^1 - I q^2 - J q^3 - K
q^4. $$ The symmetry group $Sp(1)_R \sim SU(2)_R$ is the group of
unit quaternions.

Right multiplication by $I$ on $q$ gives \bea q & \,\rightarrow\,
& q I \cr & \, \rightarrow \, & q^1 I - q^2 - q^3 K + q^4 J,\eea
which can be realized by the matrix \be\label{defs} I^R =
\pmatrix{0 & -1 & 0 & 0 \cr
            1 & 0 & 0 & 0 \cr
            0 & 0 & 0 & 1 \cr
            0 & 0 & -1 & 0 } \ee
acting on $q$ in the usual way $ q_a \, \rightarrow\, I^R_{ab} \,
q_b$. The matrices $J^R$ and $K^R$ realize right multiplication
by $J,K$ while ${\bf{1}}^R$ is the identity matrix:
\bea\label{defstwo} J^R = \pmatrix{0 & 0 & -1 & 0 \cr
            0 & 0 & 0 & -1 \cr
            1 & 0 & 0 & 0 \cr
            0 & 1 & 0 & 0 }, \qquad
             K^R = \pmatrix{0 & 0 & 0 & -1 \cr
            0 & 0 & 1 & 0 \cr
            0 & -1 & 0 & 0 \cr
            1 & 0 & 0 & 0 }. \eea
Fianlly, we define operators $s^j$ in terms of $\left\{
{\bf{1}}^R, I^R, J^R, K^R \right\}$ $$   s^1 = \pmatrix{\bf{1}^R
& 0 \cr 0 & \bf{1}^R}, \quad
                s^2 = \pmatrix{I^R & 0 \cr 0 & I^R}, \quad
                 s^3 = \pmatrix{J^R & 0 \cr 0 & J^R}, \quad
                s^4 = \pmatrix{K^R & 0 \cr 0 & K^R}.$$

We will use the $s^j$ for the quaternion basis, and write a
quaternion simply as $$ a \equiv s^j a_j $$ This facilitates both
notation, and computation. Note however that in this formalism a
quaternion is an $8 \times 8$ matrix acting to the right. If we
want it to act to the left, we need to take the transpose, which
is easily seen to correspond to quaternionic conjugation, which
we denote with a bar. As an example: $$ s^j_{ab} q_j s^k_{cd} p_k
\delta_{bd} \rightarrow q \bar{p}$$ Note that a combination like
$a \bar{a} \equiv |a|^2$ is actually a real number, multiplying
the $8 \times 8$ identity matrix.

Finally, we introduce gamma matrices
$$ \gamma^1 = \pmatrix{ 0 & 1 \cr  1 & 0 } \qquad
 \gamma^2 = \pmatrix{ 0 & I \cr  -I & 0 } \qquad
 \gamma^3 = \pmatrix{ 0 & J \cr  -J & 0 } $$
$$ \gamma^4 = \pmatrix{ 0 & K \cr -K & 0 } \qquad
 \gamma^5 = \pmatrix{ 1 & 0 \cr  0 & -1 } $$
with $1,I,J,K$ a realization of the quaternion algebra in terms of $4
\times 4$ real anti-symmetric matrices
$$ I = \pmatrix {0 & \sigma^1 \cr -\sigma^1 & 0 } \qquad
J = \pmatrix {-i \sigma^2 & 0 \cr 0 & -i\sigma^2} \qquad
K = \pmatrix {0 & \sigma^3 \cr -\sigma^3 & 0 } \qquad
$$
where $\sigma^i$ are the Pauli matrices.

\end{document}